\def\RELEASE{1}  %
\def\ANON{0}     %
\def\SQUEEZE{0}  %

\documentclass[letterpaper,twocolumn,10pt]{article}
\usepackage{usenix2020_09}
\PassOptionsToPackage{hyphens}{url}\usepackage{hyperref}

\usepackage[noend]{algpseudocode}
\usepackage{algorithmicx}
\usepackage{booktabs}
\usepackage{caption}
\usepackage{comment}
\usepackage[inline]{enumitem}
\usepackage{graphicx}
\usepackage{listings}
\usepackage{multirow}
\usepackage{xcolor}
\usepackage{xspace}
\usepackage{libertine}
\usepackage{microtype}
\usepackage{tikz}
\usepackage{tabularx}

\usepackage{subcaption}

\usepackage{inconsolata}  %

\usepackage{enumitem} %

\usepackage{makecell} %

\usepackage{relsize} %
\usepackage[most]{tcolorbox}
\tcbuselibrary{xparse,listings}
\NewTCBListing{framedlisting}{O{}}
 {%
  boxrule=0.4pt,
  colback=white,
  colframe=black,
  sharp corners,
  listing only,
  left=0mm, right=0mm, top=0mm, bottom=0mm,
  listing options={#1,frame=none, xleftmargin=0em, framexleftmargin=0em, framextopmargin=0em, belowskip=0em, aboveskip=0em},
 }

\lstdefinestyle{c++}{
    basicstyle=\linespread{0.8}\ttfamily\footnotesize\relsize{-1},
    breakatwhitespace=false,         
    breaklines=true,                 
    captionpos=b,                    
    keepspaces=true,                                     
    numbersep=5pt,                  
    showspaces=false,                
    showstringspaces=false,
    showtabs=false,                  
    tabsize=1,
    xleftmargin=1.5em,
    frame=single,
    framexleftmargin=1.5em,
}

\microtypecontext{spacing=nonfrench}

\hypersetup{
  pdflang={en},
  pdfdisplaydoctitle}

\definecolor[named]{OurPurple}{cmyk}{0.55,1,0,0.15}
\definecolor[named]{OurDarkBlue}{cmyk}{1,0.58,0,0.21}
\hypersetup{
  colorlinks,
  linkcolor=OurPurple,
  citecolor=OurPurple,
  urlcolor=OurDarkBlue,
  filecolor=OurDarkBlue}

\usepackage{titlesec} %
\titlespacing*{\paragraph}{0pt}{2mm}{2mm}
\if \SQUEEZE 1
\setlist[itemize]{
  leftmargin=*,
  itemsep=2pt,
  topsep=2pt}

\fi

\def\Snospace~{\S{}}

\newcommand{\fakepara}[1]{\paragraph{#1}}

\newcommand{\highlight}[2]{\textcolor{#1}{#2}}

\definecolor{codebg}{RGB}{255,255,255}   %
\definecolor{codeframe}{RGB}{200,200,200}
\definecolor{codestring}{RGB}{0,0,0}
\definecolor{codekeyword}{RGB}{0,0,0}
\definecolor{codecomment}{RGB}{0,128,0}
\definecolor{codeidentifier}{RGB}{0,0,0}
\definecolor{speccolor}{RGB}{0,0,180}
\definecolor{polcolor}{RGB}{160,32,240}

\lstdefinestyle{CStyle}{
    language=C,
    backgroundcolor=\color{codebg},
    basicstyle=\ttfamily\footnotesize,
    keywordstyle=\color{codekeyword}\bfseries,
    stringstyle=\color{codestring},
    commentstyle=\color{codecomment}\itshape,
    identifierstyle=\color{codeidentifier},
    showstringspaces=false,
    numbers=left,
    numberstyle=\tiny\color{black},
    numbersep=9pt,
    frame=single,
    rulecolor=\color{codeframe},
    frameround=tttt,
    tabsize=4,
    breaklines=true,
    breakatwhitespace=true,
    columns=fullflexible,
    escapechar=@,
    xleftmargin=0pt,     %
  framexleftmargin=0pt %
}
\newcommand{\evalgraphscale}{0.4}

\if \RELEASE 1
  \def\NOTES{0}
\else
  \def\NOTES{1}
\fi
\if \NOTES 1
  \newcommand{\XXX}[1]{{\color{red}{XXX {#1}}}}
  \newcommand{\antoine}[1]{{\color{teal}{[\textbf{AK:} {#1}]}}}
  \newcommand{\va}[1]{{\color{violet}{[\textbf{VA:} {#1}]}}}
  \newcommand{\todo}[1]{{\color{blue}{TODO: {#1}}}}
  \newcommand{\fake}[1]{{\color{red}{[\textbf{FAKE:} {#1}]}}}
\else
  \newcommand{\XXX}[1]{}
  \newcommand{\antoine}[1]{}
  \newcommand{\va}[1]{}
  \newcommand{\todo}[1]{}
  \newcommand{\fake}[1]{}
\fi

\newcommand{\stexttt}[1]{\texttt{{\small #1}}}
\newcommand{\eg}{e.g.\xspace}
\newcommand{\ie}{i.e.\xspace}

\if \ANON 1
  \newcommand{\sys}{PrismX\xspace}
  \newcommand{\sysx}{PrismX} %
\else
  \newcommand{\sys}{Iridescent\xspace}
  \newcommand{\sysx}{Iridescent} %
\fi

\newcommand{\codefrag}[1]{{\ttfamily\footnotesize #1}}

\begin{document}
\date{}
\title{\sys: A Framework Enabling Online System Implementation Specialization}

\if \ANON 1
  \author{Anonymous Submission \#339}
\else
\author{Vaastav Anand, Deepak Garg, Antoine Kaufmann\\Max Planck Institute for Software Systems}

\fi

\maketitle

\pagestyle{plain}

\begin{abstract}
  Specializing systems to specifics of the workload
  they serve and platform they are running on often significantly improves
  performance.
  However, specializing systems is difficult in practice because of compounding challenges: i) complexity for the developers to determine and implement optimal specialization; ii) inherent loss of generality of the resulting implementation, and iii) difficulty in identifying and implementing a single optimal specialized configuration for the messy reality of modern systems.

  To address this, we introduce \sys, a framework for automated online system specialization guided by observed overall system  performance.
  \sys lets developers specify a space of possible specialization choices, and then at runtime generates and runs different specialization choices through JIT compilation as the system runs.
  By using overall system performance metrics to guide this search, developers can use \sys to find optimal system specializations for the hardware and workload conditions at a given time.
  We demonstrate feasibility, effectivity, and ease of use.
\end{abstract} 
\section{Introduction}

Specializing system implementations to workload characteristics and hardware can
significantly improve performance and
efficiency~\cite{farshin2021packetmill,kalia:erpc,van2017automatic,miano2022domain,sriraman2018mutune,alipourfard2017cherrypick,karthikeyan2023selftune,rzadca2020autopilot,ghasemirahni2024just,wu2024tomur,ghasemirahni2022packet,deng2020redundant,sharif2018trimmer,bhatia2004automatic,molnar2016dataplane,pan2024hoda,sriraman2019softsku,dean1995selective,consel1996general,choi2025ml,srinivas2015reactive}.
Achieving these benefits requires manual modification of the system
implementations and recompilation.
Part of the performance benefit arises from
cascading compiler optimizations, e.g. by removing a feature enabling the
compiler to eliminate dead code, in turn enabling further optimizations
~\cite{wintermeyer2020p2go,panchenko2019bolt,farshin2021packetmill}.
Today, system \emph{specialization} for performance is manual, developer-driven, and
iterative.

Building optimal systems in practice is a challenge because of three
compounding factors: First, specialization is difficult for developers to
implement, as it requires trial-and-error and maintaining multiple versions of
key code paths.
Next, specialization  fundamentally comes at the cost of generality --- a
specialized system either performs poorly outside its regime,
or completely fails.
Finally, predicting performance implications of specialization choices
for different workloads and hardware is almost impossible.
To make matters worse, workload and platform conditions are dynamic
and optimal specialization choices depend on them.

We propose a fundamentally different approach:
\emph{automated online system specialization guided by observed overall system
performance}.
Our goal is to specialize performance critical system systems at runtime to the
exact momentary workload and hardware conditions, \emph{without the human
developer on the critical path}.
To enable this, we re-distribute tasks between ahead of time development and
runtime operation.
Ahead of deployment, rather than choosing a concrete set of specializations,
\emph{developers specify the space of possible specializations}, along with a
search strategy.
After deployment, our runtime iteratively \emph{generates and measures the performance of different specialized
code} versions, choosing the currently optimal specialization based on observed overall system
performance.
This allows compile-time specializations at runtime, 
suitable for search-based auto-tuning~\cite{van2017automatic,sriraman2018mutune,somashekar2024oppertune}.

To enable this, we introduce \sys, a framework for practical
development of fast and efficient systems with online specialization.
\sys enables both the incremental specialization of existing code-bases with
minimal modifications, and in-depth design for specialization for maximizing
performance.
\sys builds on LLVM and targets systems written in typical (non-managed) languages
such as C(++) or Rust that are architected around an event-loop.
Concretely, \sys requires developers to separate the code-base into a
specialized, performance critical handler core, and the remaining outer event loop code.
\sys then provides the developer with a specialization API to annotate possible
specializations in the code thereby specifying the space of possible
specializations.
Additionally, \sys provides hooks to optionally customize JIT code generation in
depth.
Finally, \sys provides a runtime API to control the exploration of different
specialized implementations as the system runs.

We integrate \sys with multiple systems and show that with
\sys enabled specializations, systems can gain a performance boost of upto 30$\times$.
Moreover, with \sys, developers can easily configure the system to automatically
explore the specialization space to adaptively find the best performing specialization
for dynamic workloads at runtime.

\section{Motivation \& Challenges}%
\label{sec:bg}

We illustrate the benefits and challenges of specialization with an example
system: a server executing square matrix multiplications for clients. To
optimize cache locality we use a blocked matrix
multiplication algorithm~\cite{carr1989blocking,wolfe1987iteration,wolfe1989more}. We
identify one key workload parameter, $N$, the matrix size, and a key
configuration parameter, $B$ the block size.

\fakepara{Code specialization improves performance.}
Both parameters offer a substantial opportunity for optimization through
specialization.
Unsurprisingly, we find fixing $B$ as a compile-time constant instead of
leaving it a variable improves throughput by up to $6.5\times$, by enabling the
compiler to unroll and vectorize the inner loops.
Note that $B$ is an internal parameter that may affect performance, but any
valid choice results in correct behavior with every workload.
Similarly, assumptions about the workload, $N$, can also simplify the algorithm,
e.g. by assuming that $N$ is a multiple of $B$ we avoid the need for copying and
0-padding partial blocks.

\begin{table}[t]%
\centering%
\small%
\begin{tabular}{lccc}
\toprule
  \textbf{Machine / Workload} & \textbf{N=1024} & \textbf{N=256}
  &\textbf{N=64} \\ 
\midrule
IceLake    & 32     & 32    & 32   \\
IvyBridge         & 16     & 16    & 4    \\
CoffeeLake       & 32     & 4     & 4    \\
AlderLake-p & 32     & 4     & 2    \\
AlderLake-e & 64     & 4     & 4    \\
\bottomrule
\end{tabular}%
\caption{Optimal value of block size, $B$, for our block matrix
  multiply, across 5 hardware platforms and 3 workloads.}%
\label{tab:mmul_hw}%
\end{table}

\fakepara{Optimal configuration depends on workload and HW.}
We now compare different block sizes for different workloads (matrix sizes) on
5 different processors.
As \autoref{tab:mmul_hw} shows, different block sizes yield optimal performance
for different workload and processor combinations.
Picking a fixed $B$ ahead of a practical deployment in a dynamic environment
will not yield optimal performance.
Moreover, even with single single-size workloads it is difficult to
predict what block size will be ideal for a concrete processor.

\subsection{Specialization is Effective but Complex}
There is a long line of prior work that has established the performance and
efficiency benefits of specializing code, and systems code more specifically.
For example, interpreted languages may generate specialized instances of
functions with constant parameters and optimize them
accordingly~\cite{de2014just,lima2020guided}.
This is a generic optimization and typically done automatically and
transparently by the runtime.
Other optimizations are specific to individual systems and concrete concerns,
such as inlining small table lookups in software network functions as if
statements~\cite{miano2022domain}.
Like most compiler optimizations, these approaches are typically guided by simple
heuristics around local metrics (e.g. frequency of the same parameter value, or
table size) based on developer intuition.

\fakepara{Specialization effects often cascade.}
A key aspect of specialization techniques is that the effects combine.
For example, PacketMill~\cite{farshin2021packetmill} first de-virtualizes
function pointers~\cite{kohler2002clickpl} for the Click modular
router~\cite{kohler:click}, and then based on this further eliminates dead code
and data structure fields.
The analysis for the latter optimization is impossible before
de-virtualization as the virtual calls prevent the compiler from analyzing the
complete packet handling code.

\subsection{Specialization is Challenging in Practice}

\fakepara{(C1) Developer Complexity.}
A key barrier to specialization is the increased complexity for developers.
Developers now need to reason about what assumptions will definitely hold and
can help specialize the system for better performance.
Next the developer needs to implement the required specialization and evaluate
it.
In practice, this frequently results in having to maintain multiple code
versions.
Unsurprisingly, this is not only complex but also laborious and frustrating,
since it is often an iterative hit-and-miss process.

\fakepara{(C2) Loss of Generality.}
Specializations may hurt performance
or break the system when the underlying assumptions are violated.
For example, a network function may process 99.999\% of packets if the processing loop
and data structures are specialized in a specific way.
However, specializing the system to this class of packets, and thereby
simplifying code and data structures may break handling of the rare 0.001\% cases.
The need to gracefully handle rare or unexpected cases forces developers to
specialize conservatively.

\fakepara{(C3) Optimal Specialization Choices.}
Modern system performance is a complex product of the emergent behaviors across
all system components, it is also not feasible for developers 
to distill this into simple local cost models.
Optimally specializing systems statically is fundamentally impossible since
concrete workload and hardware parameters affect these choices, and both
are dynamic in practice.
\section{Approach}
\label{sec:ideas}

\begin{table}[t]
\centering
\small
\begin{tabularx}{\linewidth}{lX}
\toprule
\textbf{Specialization} & \textbf{Effect} \\
\midrule
Value & Convert runtime variables to constants; triggers const. prop.,
unrolling, vectorization, .... \\
Assumption & Add boolean assumptions; triggers dead code elimination, branch
elimination, ... \\
Fast-Path & Add if-else based early exits with previously computed/cached
results. \\
Custom Code & Generate custom code at runtime. e.g.
dynamic LPM matching via nested branches. \\
\bottomrule
\end{tabularx}
\caption{\sys specialization types and their effects.}%
\label{tab:specializations}%
\end{table}

We enable practical specialization of systems for improved performance
and efficiency.
Our key idea to enable this given the challenges above, is to
\emph{automatically explore a space of possible specializations provided by the
developer at runtime based on observed overall system performance.}

\fakepara{Simple specialization with simple annotations.}
We first observe that, while developers struggle with determining \emph{optimal}
specializations and implementing a system based on them,
it is much easier for developers to suggest \emph{possible} specialization
assumptions.
We thus enable developers to annotate system code with simple annotations to
this end.
\autoref{lst:mmul-handler} shows an example of the annotations for the matrix multiplication server
above and how developers can configure specializations at runtime.
Additionally, developers can also provide LLVM transformations to perform
intrusive app-specific specializations, and thereby generate further candidate
configurations in the overall specialization space.
Crucially, we do not rely on the developer to filter or rank specializations.
\autoref{tab:specializations} shows the broad class of specializations supported
by \sys and their corresponding effects.

\fakepara{Guarded specialization with a fall-back.}
Later, when automatically instantiating specialized system versions through JIT
compilation, we inject guard conditions into the code to use the specialized
version of the code when applicable. If the specialized version is not applicable,
the guard conditions ensure that we fall back to the
unspecialized version of the system code for that invocation.
While triggering this guard incurs overhead, it does ensure correctness.
If the guard triggers sufficiently rarely, in combination with the
benefit for the common case, the specialization is still a net win.

\fakepara{Online exploration guided by developer policy.}
Finally, we completely forgo cost models and heuristics for predicting which
choice in the specialization space is optimal, and instead explore different
points online as the system runs.
Our guards ensure that the system always behaves correctly in this process.
We rely on the developer to provide a policy for guiding this exploration.
The developers control when to explore which points, manually, through
existing auto-tuning solutions or by leveraging simple search strategies from
our library.
A key task of this policy is comparing the (application specific) overall system
performance metrics, e.g. request throughput, when running different
specialization points.
Overall system performance metrics also implicitly factor in any overheads, e.g.
triggering specialization guards for some calls.
This ensures that the system converges to the optimal specialization point for
the concrete combination of dynamic system, workload, and hardware conditions.
\section{\sys Design}%
\label{sec:design}

We now present the design of our approach in \sys.

\if \ANON 1
\begin{figure}[t]%
\centering%
\includegraphics[width=\linewidth]{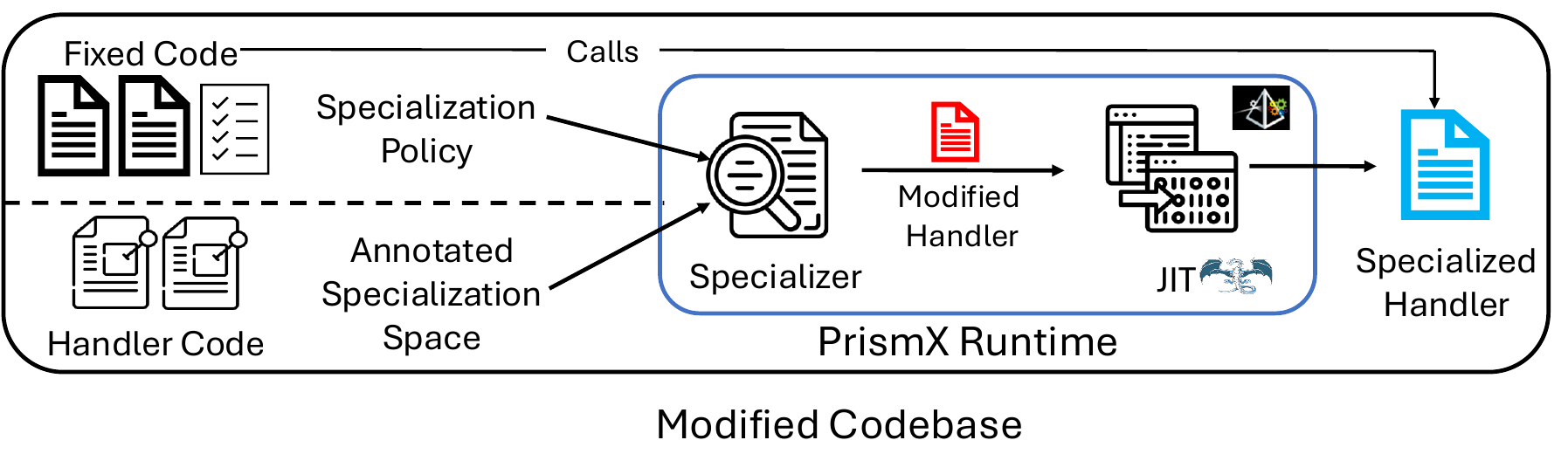}%
\caption{Architecture of a \sys-specialized system.}%
\label{fig:sys_design}%
\end{figure}
\else
\begin{figure}[t]%
\centering%
\includegraphics[width=\linewidth]{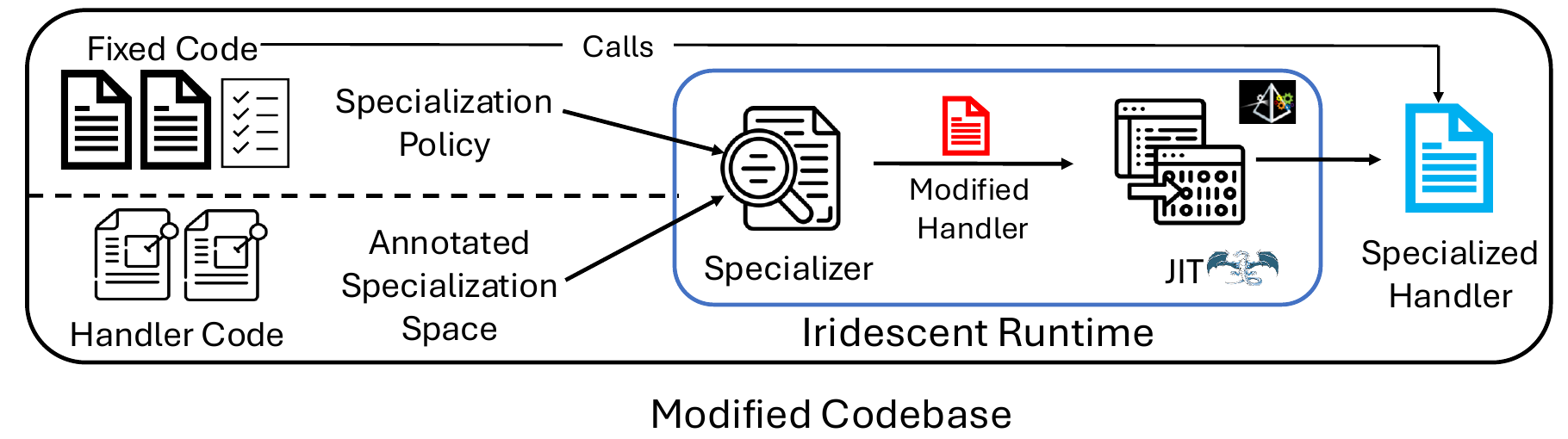}%
\caption{Architecture of a \sys-specialized system.}%
\label{fig:sys_design}%
\end{figure}
\fi

\begin{figure}[t]%
\begin{subfigure}{\linewidth}%
\begin{lstlisting}[style=CStyle]
void matmul(u64 *L, u64 *R, u64 *O, int N, int B) {
  B = @\highlight{speccolor}{spec\_enum("B",B,2,4,8,16,32,64);}@
  N = @\highlight{speccolor}{spec\_general("N",N);}@
  for (int kk = 0; kk < N; kk += B)
    for (int jj = 0; jj < N; jj += B)
      for (int i = 0; i < N; i++)
        for (int jjj = jj; jjj < jj + B; jjj++)
          for (int kkk = kk; kkk < kk + B; kkk++)
            M3[i*N+jjj] += L[i*N+kkk] * R[kkk*N+jjj];
}
\end{lstlisting}%
\subcaption{Handler Code}%
\label{lst:mmul-handler}%
\end{subfigure}%

\begin{subfigure}{\linewidth}%
\begin{lstlisting}[style=CStyle]
@\highlight{polcolor}{\sysx}@ @\highlight{polcolor}{rt("handler\_code.ll");}@
void spec_policy() {
  Conf N_cs[2] ={{}, {{"N", 256}}};
  do { for (c : @\highlight{polcolor}{cartesian(rt->spec\_space(), N\_cs)}@) {
          @\highlight{polcolor}{rt.specialize(c)}@; tput_counter = 0;
          sleep(@$\hdots$@); // Sleep for some time to monitor
          if (tput_counter > best) {
            best = tput_counter; best_c = c; } }
    @\highlight{polcolor}{rt.specialize(best\_c);}@
  } while(every 1min);
}
int main() {
  auto *matmul = @\highlight{polcolor}{rt.handler("matmul")};@
  @\highlight{polcolor}{rt.reg\_opt\_pipeline(opt\_passes);}@
  launch_thread(spec_policy);
  while(true) { // Main processing loop
    Req *req = get_req(); Resp *res = new Resp();
    matmul(req.A, req.B, res.C, 2, req.N);
    tput_counter++;
  }
}
\end{lstlisting}%
\subcaption{Fixed Code}%
\label{lst:mmul-fixed}%
\end{subfigure}\hfill%
\caption{MMulBlockBench microbenchmark}%
\label{lst:mmul}%
\end{figure}

\begin{table}[t]
\centering
\begin{tabularx}{\linewidth}{lX}
\toprule
\multicolumn{2}{l}{\textbf{\sys Specialization API}} \\
\midrule
\codefrag{spec\_enum(lbl, x, ...)} & Enumeration spec. point (x$\in$...)\\
\codefrag{spec\_range(lbl, x, l, h)} & Range spec. point (l$\le$x$\le$h)\\
\codefrag{spec\_generic(lbl, x)} & Policy-controlled spec. point \\
\codefrag{spec\_assume(lbl, cond)} & Specialization assumption \\
\codefrag{spec\_custom\_*(lbl, ...)} & Custom spec. point \\
\midrule
\multicolumn{2}{l}{\textbf{\sys Policy API}} \\
\midrule
\codefrag{\sysx(handler\_ir)}              & Initialize runtime \\
\codefrag{.handler(h)}          & Get specialized handler \\
\codefrag{.spec\_space()}       & Obtain code spec. space  \\
\codefrag{.specialize(c)}            & Choose specialization; \codefrag{c} maps
spec. point labels $\mapsto$ values \\
\codefrag{.add\_custom\_spec(n, gen)} & Add app specialization\\
\codefrag{.customize\_opts(passes)} & Modify codegen optimizations \\
\bottomrule
\end{tabularx}%
\vspace{1mm}%
\caption{\sys API overview.}%
\label{tab:hooks}%
\vspace{-3mm}%
\end{table}

\subsection{Anatomy of a \sys System}

\fakepara{System Requirements for \sys.} 
\sys targets performance critical systems not implemented in managed
languages. We observe that typical performance critical systems code is almost
exclusively architected with an event loop structure -- an outer loop handling a sequence of events, such
as processing requests or data elements, and an inner handler that processes the event.
While \sys is designed to specialize these systems in varying dynamic environments, 
for these systems to fully gain the benefits of \sys, these systems should
have stability in the workload and operating conditions
for a sufficient period of time.
Conditions should at least remain
stable for the time \sys takes to re-compile a new specialized variant
(typically 10--400\,ms, see \autoref{sec:cost}).

\fakepara{Isolating performance critical code for specialization.} 
Since \sys targets performance critical systems not implemented in managed
languages, modifying code of the running system is non-trivial in general.
We leverage the event-loop structure for pragmatic specialization, by requiring developers
to separate the system code into two parts (\autoref{fig:sys_design}):
(i) the \emph{fixed code}, including initialization and the outer loop;
(ii) the performance-critical event \emph{handler code}.
The developer compiles the fixed code as before, but integrates calls to the
\sys API (\autoref{tab:hooks}).
In the handler code, the developer adds lightweight \sys specialization API
calls and compiles to LLVM intermediate representation.
In the fixed code, the developer obtains function pointers for the handlers
through the \sys policy API, and calls these as before.
\autoref{lst:mmul} illustrates this with our matrix multiplication server running example.

\fakepara{Handling Global State.} During dynamic code generation, \sys will link handlers against symbols in the
fixed code, to enable calls and accesses to global state.
Since \sys repeatedly rebuilds handlers, \emph{\sys requires that the handler
code does not include definitions of global variables that must be preserved.}
The developer has to move all global state to the fixed code, and reference it from
the handlers.

\subsection{Defining Specialization Space} 

The developer defines the specialization space through \emph{specialization
point} annotations in the handler code.
At runtime, \sys replaces these annotations with specialized code,
instrumentation, or disables them, guided by the policy. 
\sys offers four types of specialization points:

\fakepara{Value Specialization Point.}
A value specialization point signals to \sys that the handler code could be specialized
to a constant for the wrapped expression.
\codefrag{spec\_enum} indicates the value could be one of the specified values;
\codefrag{spec\_range} instead specifies a range of integers;
\codefrag{spec\_generic} leaves it to the policy to determine possible values.

\fakepara{Assumption Specialization Point.}
This provides a \emph{possible specialization assumption} to \sys.
The condition is a boolean predicate that \sys can provide to compiler
optimizations to further optimize, through LLVM's builtin \codefrag{llvm.assume}.
Crucially, the assumption need not always hold. Unlike for the LLVM intrinsic
where incorrect behavior results, \sys catches this with
its specialization guards.

\fakepara{Fast-Path Specialization Point.}
This is a generic version of the hot key specialization of
Morpheus~\cite{miano2022domain}. It works in two phases: (i) the instrumentation
phase: \sys inserts instrumentation code that samples (sampling rate selected by
the developer) invocations of the specialized function to find the most popular
inputs to the function along with their calculated outputs; (ii) the
specialization phase: \sys updates the specialized function to have a
specialized if-else chain where the top-N inputs along with their outputs are
converted into a series of if-else checks to avoid paying the cost of
re-computation for the heavy-hitting inputs. The value of N may be
developer-provided or could be configured as a runtime constant that can be
further specialized. If the input does not match any of the branches in the
if-else chain, the computation simply defaults to the generic version.

\fakepara{Custom User-Defined Specialization Point.}
These specialization points, annotated through \codefrag{spec\_custom\_*},
invoke developer-defined specializations registered in the fixed code through
\codefrag{add\_custom\_spec}.
An example is generating an if-else chain as a special-case fast-path for an
expensive data structure lookup or computation (\autoref{ssec:case:liblpm}).

\subsection{Defining Specialization Policies}
The developer implements the specialization policy in the system's fixed code
using the \sys policy API.
The policy decides what, when, and how \sys should specialize.

\autoref{lst:mmul-fixed} illustrates how 
developers can control the exploration and selection of specializations.
In line 4, the \stexttt{spec\_space} API returns the specialization space
generated by the annotations in the handler code.
We combine this space with other specializations through a
Cartesian product to obtain a complete set of specializations.
In lines 4--8, the fixed code automatically tries out the different
combinations in the set of specializations and chooses the best
performing combination.
To choose the best, the specialization policy uses the target end-to-end performance metric, in this case the throughput.
We trigger re-exploration at a fixed-time interval to adapt to
workload changes.

The developer can also optionally configure custom optimization passes through
the \codefrag{customize\_opts} API to interpose on and customize code generation.
Developers may also choose to enable instrumentation for select specialization points to
dynamically identify opportunities for specialization.
For example, the policy for the matrix multiplication server could detect the
most frequent matrix size $N$, and only decide to specialize if
$\ge$70\% of requests have this value.

\fakepara{Exploring the specialization space.} \sys provides a periodic exhaustive search strategy for simple cases as a
library routine, with a simple call-back for providing the strategy with the
system performance metric.
For more complex cases, \sys integrates with OPPerTune~\cite{somashekar2024oppertune}
to leverage its state-of-the-art search strategy for configuration tuning of web services.
Additionally, \sys also provides a thin interface for developing custom search strategies.
We expect most systems will either utilize existing auto-tuning or implement custom strategies
maximizing overall system performance, using the \sys building blocks.

\subsection{Specialization Runtime}

The specialization runtime has two key components: (i) the specializer generating code by applying selected specializations;
(ii) the JIT compiling and optimizing the modified specialized code and making
it available to the fixed code.

\subsubsection{Specializer}

The specializer generates the specialized version of the
handler code.
It operates at function granularity, and generates versions of functions with
specialization points replaced depending on the chosen specialization configuration.

For each specialization point, \sys performs the following actions,
depending on the configuration the policy supplied for this point.
If the policy marks a point as disabled, the \sys specializer simply removes
the annotation and replaces it with the original value, or skips over the point if it is an assumption point.
To specialize an enabled value specialization point, \sys replaces the specialization
point annotation in the handler code with the constant value supplied by the
policy.
By default, the specializer will also insert a specialization guard, which the developers may explicitly disable.
For a custom specialization point, the specializer replaces the
specialization point with the custom source code provided by the previously
registered handler for that point type.
The recompilation process then enables standard compiler optimization
to take advantage of the newly inserted constants or code.

\fakepara{Instrumentation.} Some specialization policies benefit from collecting
runtime data to generate possible specialization configurations. To support such
data collection, the specializer can optionally enable instrumentation for each specialization
point. For a specialization point with instrumentation enabled, \sys
additionally generates code for collecting and storing the observed actual
values.
The policy retrieves this information included in the result of the
\codefrag{spec\_space} call.

\subsubsection{JIT}

\fakepara{Compiling the specialized code.} The specialization runtime adds the specialized code generated by the specializer
to the JIT. The JIT compiles the code and runs the developer-specified pipeline of optimization passes,
including any custom developer-provided optimization passes, to generate an optimized version of the specialized code.
By default, the JIT applies the default \texttt{O3} optimization pipeline if the developer has not
provided a specific pipeline. In addition, the JIT also keeps a copy of the original, non-specialized version of
the specialized code. This generic code serves as fallback for situations where
the specialized code is not applicable.

\fakepara{Using the specialized code}. To use the specialized handlers, the
fixed code obtains function pointers to specialized code from the specialization runtime. The fixed
code then uses these pointers to invoke the specialized handlers. If no specialization has been
enabled the handlers default to the non-specialized versions of the
functions. 
For simplicity, the fixed code only does this once at start of
execution.
The JIT creates a trampoline function which calls the most recent specialized
version of the function. The trampoline function is stored at a fixed address and does
not change across specialization iterations.

\subsubsection{Correctness through Specialization Guards}

\sys inserts a runtime check called a specialization guard for each
enabled
specialization point. The policy can disable this guard for individual
points. The
specialization guard ensures that the specialization conditions for generated 
code holds during execution. For this, the specializer inserts an early exit from the specialized version of the code on guard
failure by throwing an exception. The \sys trampoline function catches these
exceptions and re-routes control to the original non-specialized
version of the function transparently without exposing the exception to the
fixed code.

\fakepara{Restoring state and side effects.}To ensure correct restoration of the state, the specializer will call a developer-defined
cleanup function for reversing any side-effects before throwing the exception.
It is critical to note here that not all side-effects are reversible (\eg sending a packet to a neighbor), so \sys
leaves the developer with the necessary control.

\subsection{Prototype Implementation}

We have implemented \sys in $5.5K$ lines of C++.
Our implementation uses the LLVM IR and JIT~\cite{lattner:lllvm}
to generate the specialized code at runtime. The specializer is implemented as a set of LLVM transformation passes 
that operate on the LLVM IR of the instrumented handler code.
\section{Case Studies}
\label{sec:case}

We evaluate \sys with four open-source systems and libraries and the MMulBlockBench microbenchmark.
We explain each system along with the changes we made for integrating \sys below.

\subsection{Tiling Benchmarks}

\fakepara{MMulBlockBench.} We implement a blocked/tiled version of matrix multiplication algorithm~\cite{carr1989blocking,wolfe1987iteration,wolfe1989more}.
To integrate this benchmark with \sys, we convert the tile size, $B$, as a value specialization point.

\fakepara{CloverLeaf.} We use a C-version of the CloverLeaf application from the SPEC ACCEL2023 benchmark~\cite{specaccel}. In this application,
we implement tiled-versions of the various mathematical kernels, and use value
specialization points for the tile sizes, $T_i$ and $t_J$.
In total, this application has a 24 unique specialization points, each of which that can take 7 unique valeus resulting in a total of $7^{24}$ possible
configurations.

\subsection{LibLPM}
\label{ssec:case:liblpm}
LibLPM~\cite{liblpm} is an open-source Longest Prefix Match (LPM) library written in C with built-in support for IPv4 and IPv6 addresses.
LPM is a key operation in packet routing for finding the closest matching route in routing tables
for incoming packets.
To integrate LibLPM with \sys, we create two different specializations:

\fakepara{LibLPM-FP} specializes the lookup function to add a fast-path specialization point. This specialization
adds an exact match if-else fast-path for the top-N input IP addresses.

\fakepara{LibLPM-NI} creates a code-generation specialization point in the LPM library. The code-generation specialization point generates
a new version of the lookup function which generates a nested-if-else tree of checks consisting of prefix match checks for the incoming
address for each lpm entry. The nested-if-else tree starts at the least specific matching rule (based on prefix length)
and progressively checks for the most specific matching rule until it can't find one anymore. With this code generation specialization,
we embed the prefix rules directly into the codebase,
forgoing expensive dynamic datastructure traversals and enabling further
optimizations for individual branches.

\fakepara{LibLPM-NI-FP} combines the previous two specializations by adding a fast-path specialization point in the generated nested-if specialized
lookup function.

\subsection{TAS}

TAS~\cite{kaufmann:tas}, TCP acceleration as a service, is a lightweight software TCP network fast-path that is optimized for common-case client-server RPCs. TAS executes common-case TCP operations in an isolated fast path that uses DPDK~\cite{software:dpdk}, while handling corner cases in a slow path.

To integrate TAS with \sys, we convert its \texttt{BATCH\_SIZE} macro to a value specialization point. In TAS, the \texttt{BATCH\_SIZE} variable is used in three different scenarios: (i) to determine the number of packets to read from the NIC, (ii) to determine the number of packets to read from the application queues, and (iii) to determine the number of packets to read from the queue manager. For more fine-grained control, we convert each of these usage instances of the \texttt{BATCH\_SIZE} to three separate runtime constant specialization points - (i) \texttt{rx\_batch}, (ii) \texttt{queues\_batch}, and (iii) \texttt{qman\_batch}.

\subsection{FastClick}

FastClick~\cite{barbette2015fast} is an extended version of the Click Modular Router~\cite{kohler:click} with Netmap and DPDK support for running the modular router in userspace. 
A FastClick (or Click) router is assembled from individual packet processing modules called elements. Each element implements simple router functions such as packet classification, scheduling, routing, and interacting with network devices.
Users configure the router with different graphs of connected elements.

To integrate FastClick with \sys, we modify the \texttt{LinearIPLookup} element of FastClick to create a \sys-enabled \texttt{LinearIPLookup} element. 
The original element uses a linear search algorithm to find the best matching route for an incoming packet based on the routing table. 
We modify the packet-processing function of the \texttt{LinearIPLookup} element to add a fast-path specialization point.
The target of this specialization point is the internal lookup function of the \texttt{LinearIPLookup} element.

\subsection{Network Functions}

Network Functions is a suite of the network functions that include DPDK and ebpf implementations of common network functions such as a NAT, Router, policer, among others. These functions have been developed and used for evaluation by various research projects such as Pix~\cite{iyer2022performance} and Vigor~\cite{zaostrovnykh2019verifying}. We extract the implementations from the Pix artifact~\cite{pixartifact}.

To integrate the Network Functions with \sys, we convert the \texttt{BATCH\_SIZE} used by the DPDK version of these functions into a runtime constant specialization point. In the network functions, the {BATCH\_SIZE} controls the number of packets that are read from a network port at a given time.
\section{Evaluation}%
\label{sec:eval}

In this section, we evaluate how much \sys improves performance of systems by enabling online specializations. We show how we can use online specialization with \sys for three use-cases: (i) enabling compile-time optimizations at runtime;
(ii) enabling incremental specializations at runtime; (iii) design exploration at runtime. We answer the following questions:

\begin{itemize}[itemsep=0pt]
  \item Do \sys-enabled optimizations improve performance? (\autoref{sec:eval_opt}) 
  \item Can \sys find an optimal specialization through exploration at runtime under dynamic conditions? (\autoref{sec:eval_exploration})
  \item What is the overhead of using \sys? (\autoref{sec:cost})
\end{itemize}

\subsection{Experimental Setup}

\fakepara{Testbed.} For our experiments, unless otherwise stated, we configure two identical machines as client
and server. They are directly connected with a pair of 100\,Gbps
Mellanox ConnectX-5 Ethernet adapters. Each machine
has two Intel Xeon Gold 6152 processors at 2.1\,GHz, each
with 22 cores for a total of 44 cores and 187\,GB of RAM per
machine. We run Linux kernel 5.15 with Debian 11.

\subsection{Compiler Optimizations at Runtime}
\label{sec:eval_opt}

\begin{table}[t]
\centering
\small
\begin{tabular}{lccc}
\toprule
\textbf{Machine}          & \textbf{Constant} (c) & \textbf{Variable} (v) & \textbf{Benefit} \\ 
(Processor)      & (cycles/op)  & (cycles/op) & (v/c)\\
\midrule
IceLake    & 175297    & 284944             & 61\% \\ 
IvyBridge         & 250434    &   661295           & 246\% \\ 
CoffeeLake       & 168817    &  581130             & 348\% \\ 
AlderLake-p & 173350    & 583724             & 336\% \\
AlderLake-e & 206924    & 557572           & 269\% \\
\bottomrule
\end{tabular}
\caption{Impact of turning block size, $B$, as compile-time constant at runtime for N=64 for different hardwares (\autoref{tab:mmul_hw})}
\label{tab:mmul_compilation}
\vspace{-3mm}
\end{table}

\begin{figure}[t]%
\centering%
\includegraphics[scale=\evalgraphscale]{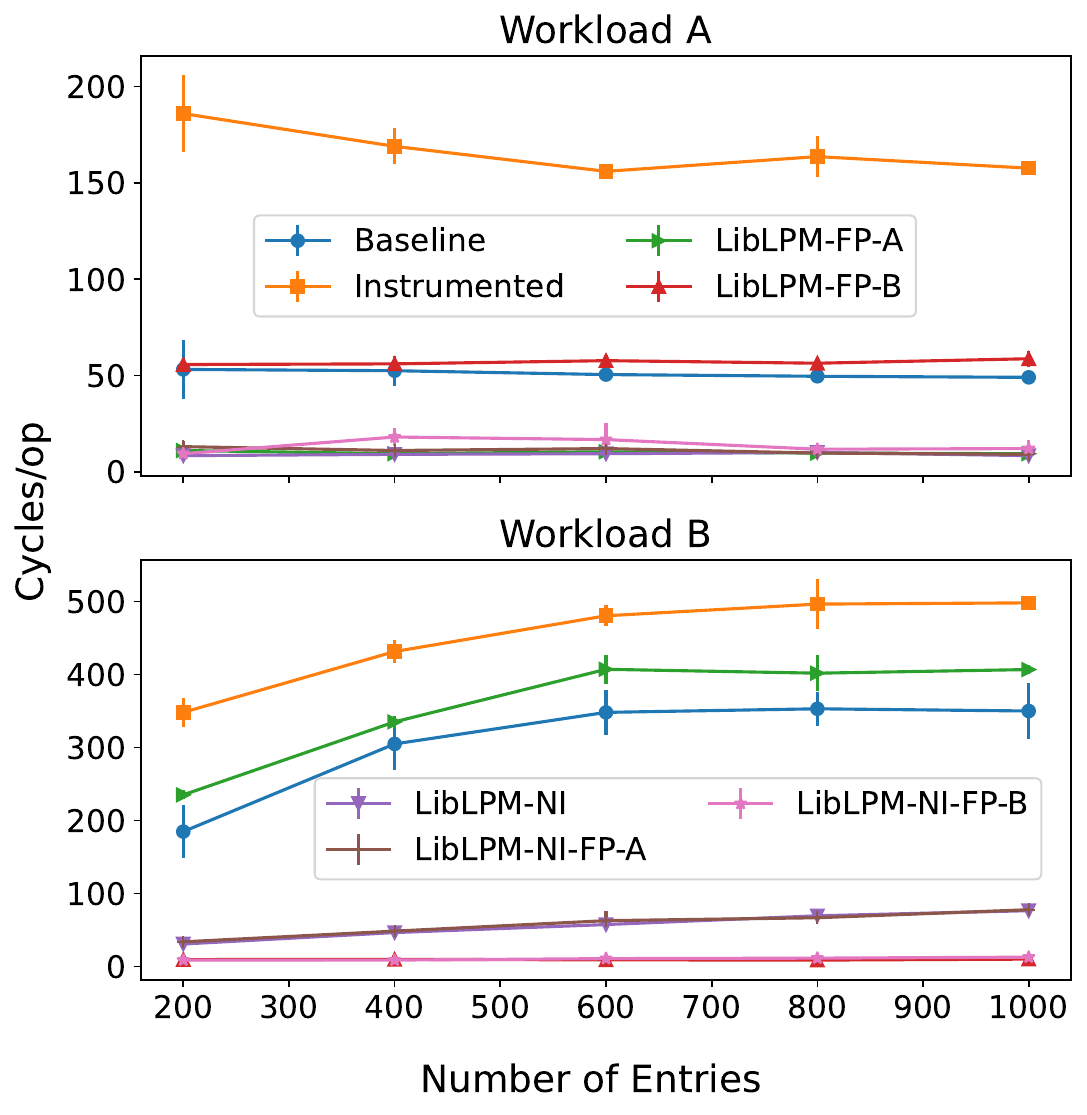}%
\caption{Average cycles required to perform a table lookup with varying table
sizes under two different workloads.}%
\label{fig:lpm_execution}%
\end{figure}

Compile-time optimizations often produce more performant code.
These compile time optimizations include
constant propagation, loop unrolling, dead-code elimination, and use of vectorization instructions.
With \sys, we can enable compile time optimizations using runtime data. 

\fakepara{\sys enables cascading compiler optimizations.} First, we show the potential benefits of enabling compiler optimizations at runtime for the MMulBlockBench microbenchmark 
in \autoref{tab:mmul_hw}. Specifically, for the workload $N=64$, we compare the performance of keeping the optimal block size, $B$,
as a runtime parameter to that of converting it into a constant at runtime using \sys, thus enabling downstream compiler optimizations.
For both configurations,
we execute the function a fixed number of times and measure the number of cycles spent per function execution.
\autoref{tab:mmul_compilation} shows the benefit of converting $B$ into a compile-time constant for different machines.
For each machines, we get at least 50\% reduction in cycles, and greater than 240\% reduction in cycles for four of the five operating conditions. 
This improvement is a direct impact of \sys's specialization allowing different compiler optimizations to cascade and combine together
to produce a more efficient version of the code. In this specific case, the specialization of $B$ allows the JIT compiler
to first easily unroll the loops in the matrix multiplication function and then leverage optimized vector instructions. 

\begin{figure}[t]%
\centering%
\includegraphics[scale=\evalgraphscale]{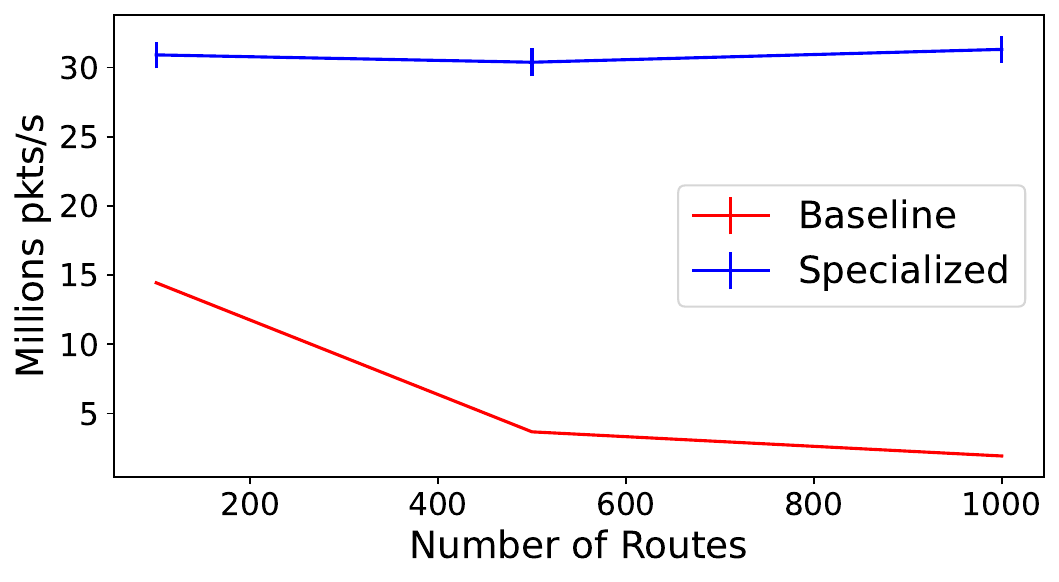}%
\caption{FastPath router throughput with \sys specialized
  LinearIPLookup element for different routing table sizes at 100\%
  fast-path hit rate.}%
\label{fig:fastclick_nroutes}%
\end{figure}
 
\begin{figure}[t]%
\centering%
\includegraphics[scale=\evalgraphscale]{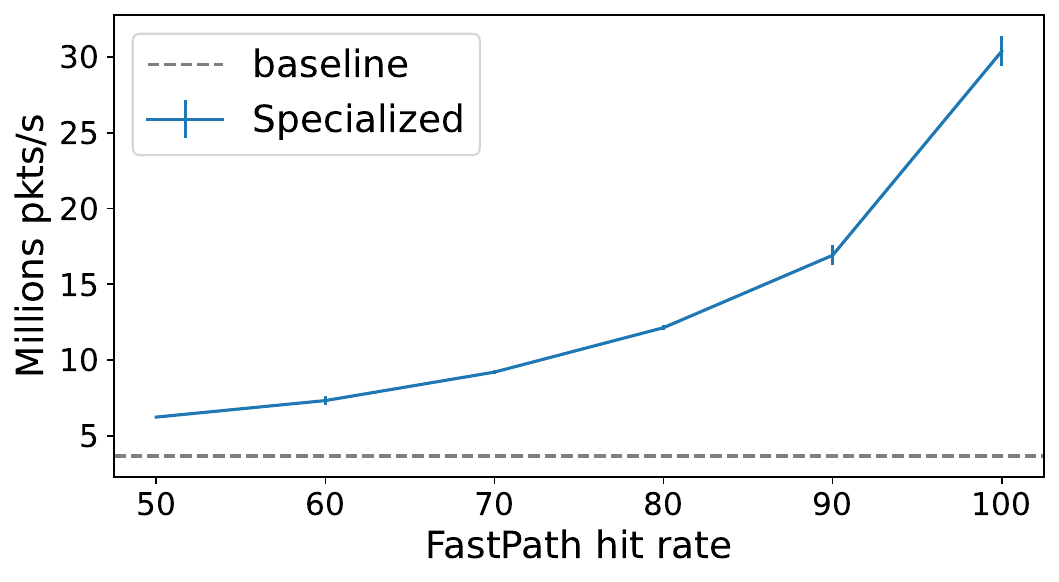}%
\caption{FastPath router throughput with \sys specialized
  LinearIPLookup element for varying fast-path hit rates.}%
\label{fig:fastclick_hr}%
\end{figure}

\fakepara{\sys incorporates existing specializations.} To showcase \sys's ability to easily incorporate existing runtime specializations, we implement Morpheus'~\cite{miano2022domain} map hot-keys specialization of  as a fast-path specialization point in the packet processing function of FastClick's LinearIPLookup element.
We configure the fast-path specialization point with a 0.01\% sampling rate for the instrumentation phase. 
We execute the FastPath router on one machine and treat it as the device under test and execute Pktgen to execute an open-loop workload to generate packets with destination ip addresses from a given set of IP addresses.
We repeat this experiment for different sizes of the routing table in the
LinearIPLookup as well as different sizes of the fast-path to measure the
throughput of the router under different conditions.
\autoref{fig:fastclick_nroutes} shows that \sys achieves  up to 15$\times$
improvement in the throughput based on the size of the table at 100\% fast-path
hit rate. The improvement increases with larger table sizes, as the basline
requires longer linear scans here.
\autoref{fig:fastclick_hr} shows that for a router with a 1000-entry routing
table, even with a 50\% fast-path hit rate, \sys achieves an approximately
5$\times$ throughput improvement. The improvement increases with the
the fast-path hit rate.

\fakepara{\sys enables custom specializations.} Next, we show the benefits of custom compile-time optimizations for LibLPM. 
For this experiment, we set up eleven different configurations resulting from a combination of five different LPM table sizes and two different workloads - Workload A and Workload B. In Workload A, the incoming IP address is an IP address that matches a very specific prefix entry in the routing table.
In Workload B, the incoming IP address is an IP address that only matches the LPM prefix entry of prefix length 0. Thus, Workload A and Workload B
cover the best and worst cases respectively for the LPM lookup function. 
We execute these workloads with seven specializations:
\begin{enumerate}[label=(\roman*),nosep, leftmargin=*, itemsep=0pt, topsep=0pt]
  \item Baseline: no \sys-enabled specialization.
  \item Instrumented: \sys-enabled specialization capturing the
    most popular inputs (10\% sampling); data used to guide the fast path
    specializations.
  \item LibLPM-FP-A: fast-path specialization of
    size 1 specialized for Workload A.
  \item LibLPM-FP-B: fast-path specialization of size 1 specialized
    for Workload B.
  \item LibLPM-NI: nested-if code generation specialization.
  \item LibLPM-NI-FP-A: Combine LibLPM-NI and -FP-A.
  \item LibLPM-NI-FP-B: Combine LibLPM-NI and -FP-B.
\end{enumerate}

\autoref{fig:lpm_execution} shows average cycles per LPM lookup
lookup function for all different combinations of configurations and
specializations. For Workload A, all three specializations,
LibLPM-FP-A, LibLPM-NI, LibLPM-NI-FP-A, achieve up to $6\times$
reduction in cycles per execution.
However, unlike the LibLPM-FP and LibLPM-NI-FP-A specialization, this LibLPM-NI specialization does not require an additional instrumentation phase to sample executions to find the optimal candidates for the fast path. LibLPM-NI improves performance across the board for all workloads by up to $6\times$.
Despite this performance boost, LibLPM-NI is not the best specialization for Workload B. For Workload B, the best optimizations are
the *-FP-B specializations as they improve by up to $30\times$
as all instantiations of Workload B hit the fast-path.

\subsection{Runtime Design Exploration}
\label{sec:eval_exploration}

\begin{figure}[t]%
\centering%
\includegraphics[scale=\evalgraphscale]{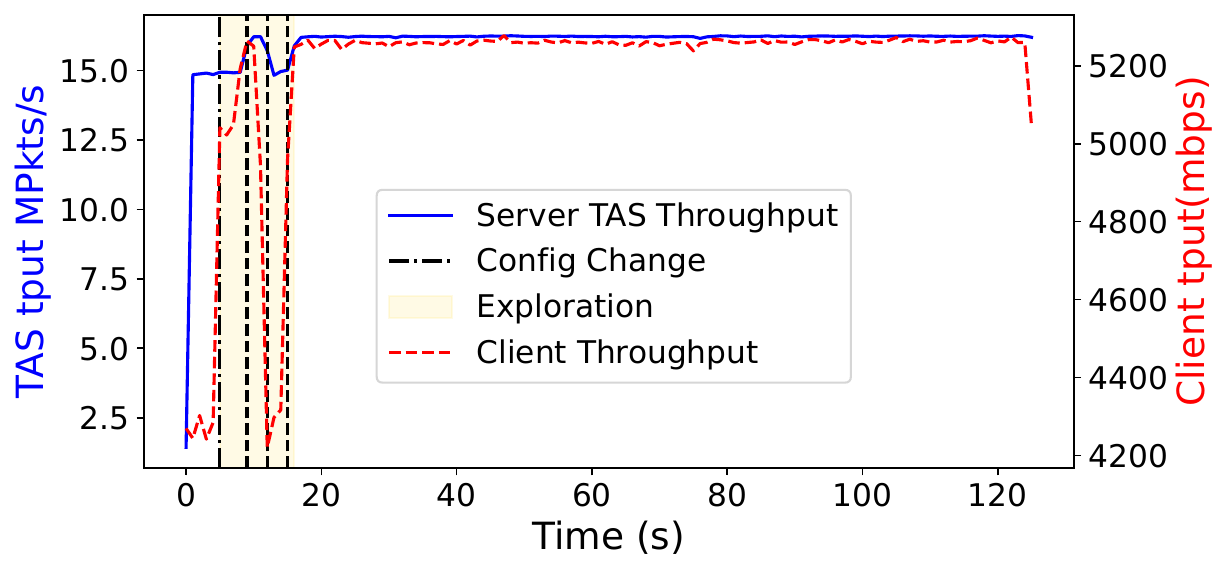}%
\caption{Automatic exploration and specialization of different
  configurations for TAS.}%
\label{fig:tas_adaptation}%
\end{figure}

\fakepara{\sys enables runtime design exploration.} We show runtime design exploration for TAS~\cite{kaufmann:tas}.
We experiment in a typical server-client setting, with the server
providing an echo service. The client is multi-threaded and executes an open-loop
workload of 64 byte packets. Both the server and client are TAS-enabled.
In our experiment, we modify the server-side TAS and measure the throughput of the
server side TAS as millions of packets processed per second. 
With \sys we explore different values of the \texttt{rx\_batch} in the TAS source code on the server side.
\autoref{fig:tas_adaptation} shows \sys automatically determining and
then selecting the best-configuration on the server side by exploring
the different values of the batch size for different code locations.

\begin{figure}[t]%
\centering%
\includegraphics[scale=\evalgraphscale]{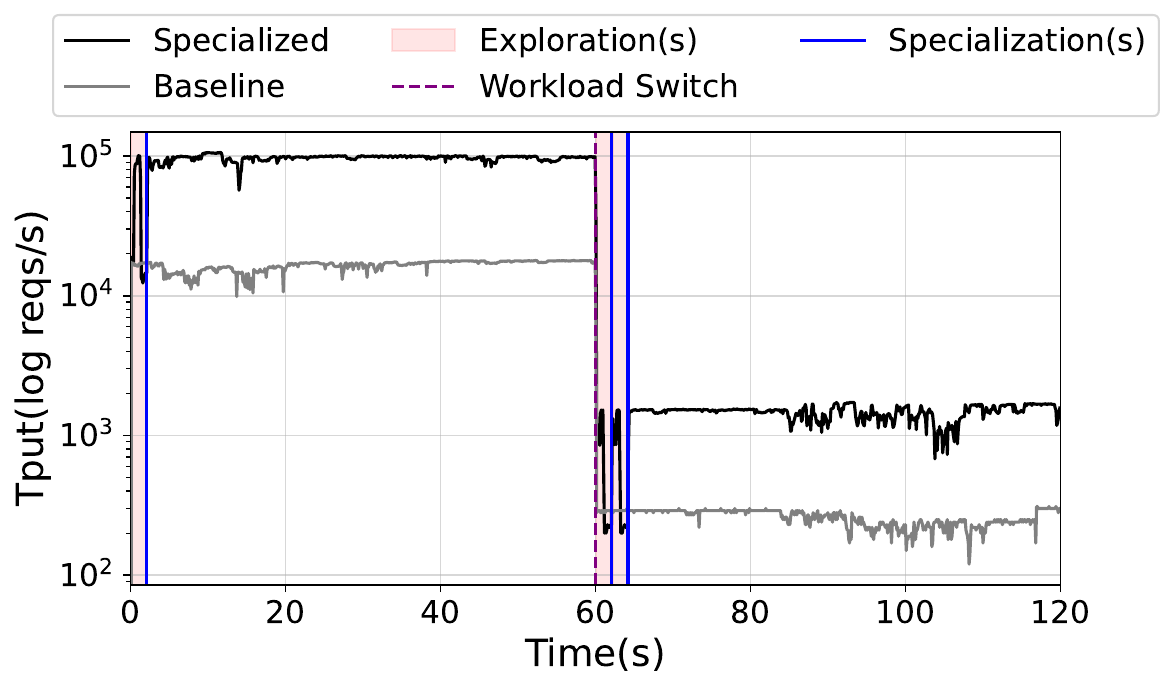}%
\caption{Automatic exploration and specialization of MMulBlockBench
  for two different workloads.}%
\label{fig:mmul_adaptation}%
\end{figure}

\fakepara{\sys automatically adapts to changing workloads.} To show that \sys can automatically adapt to workloads, we first use the MMulBlockBench microbenchmark.
We execute a \sys-enabled version of the function and a non-\sys enabled baseline.
In the baseline, the block size $B$ is a runtime parameter. In the \sys-enabled version, \sys explores different
values of $B$ by specializing the value in the code and converting $B$ into a compile-time constant.
We execute each version with two different workloads in succession, each lasting one minute.
We measure the overall throughput (executions/s) for both configurations.
\autoref{fig:mmul_adaptation} shows how
\sys explores the different block sizes and finds the most
performant configuration compared to the baseline.
Moreover, \sys can automatically detect the workload change based on the
drop in throughput and restart the A/B testing process.

\begin{figure*}[t]%
\centering%
\begin{subfigure}{\linewidth}%
\includegraphics[width=\linewidth]{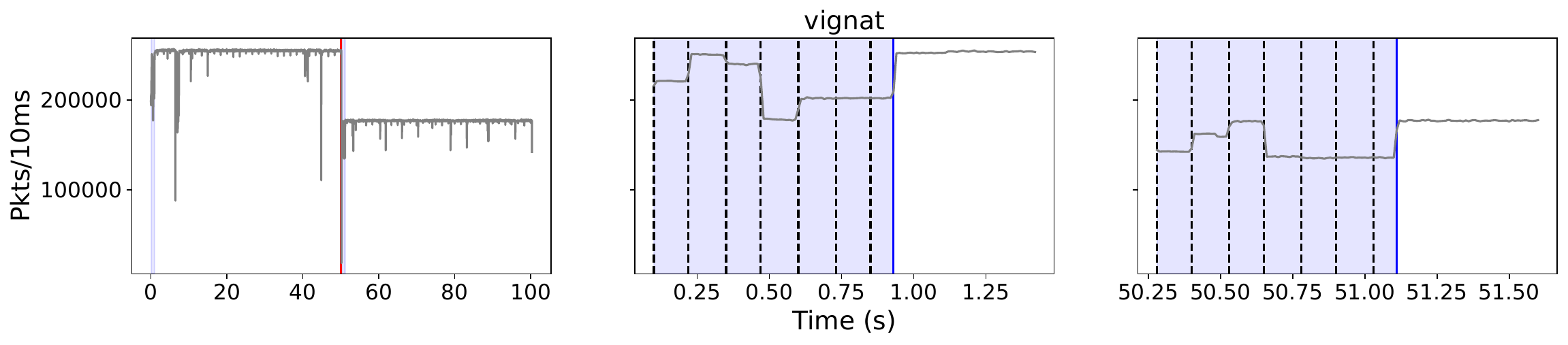}%
\end{subfigure}%

\begin{subfigure}{\linewidth}%
\includegraphics[width=\linewidth]{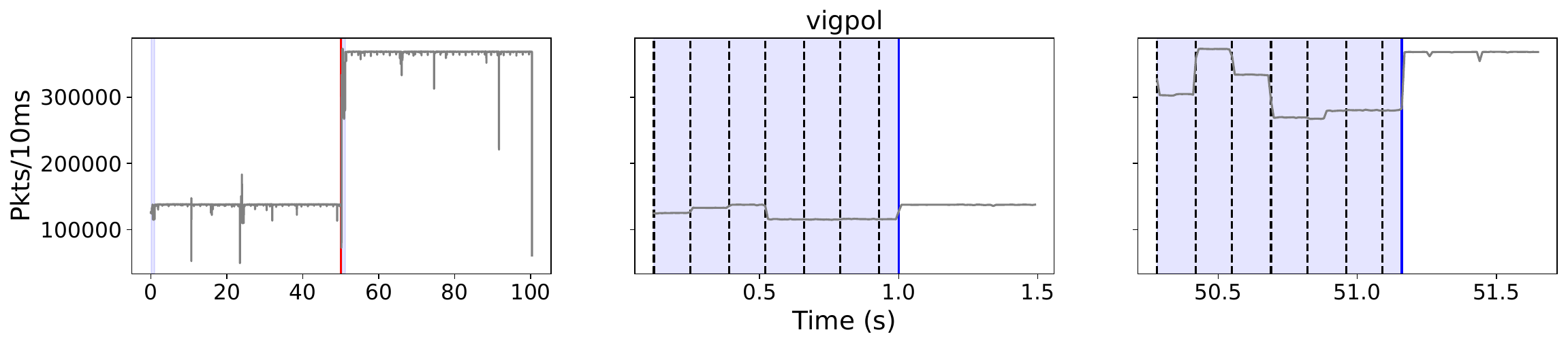}%
\end{subfigure}%
\caption{Batch size exploration with Network Functions}
\label{fig:network_adaptation}
\end{figure*}

Next, we show runtime design exploration and adaptation for two different network functions, vignat and vigpol, from vigor~\cite{zaostrovnykh2019verifying}.
We setup our experiment with a packet generator on one machine and a device under test (DUT) on the other machine.
We use two ports for the DUT. The first port acts as the internal facing port whereas the second port as the external facing port.
The DUT executes the specific network function for every incoming packet generated by the packet generator. The network function implementation has different execution pathways for packets depending on
whether they arrive on the internal port or the external port.
The network function is enabled with the \sys specialization policy to trigger an exploration phase to find the best performing \texttt{BATCH\_SIZE} whenever there is a change in the workload.
We execute the experiment in two phases: in the first phase, packets arrive only on the external facing port and in the second phase, packets arrive only on the internal facing port.
\autoref{fig:network_adaptation} shows the results for the NAT and the Policer network functions. \sys finds the best performing configuration in each phase with different optimal values.

\begin{figure}[t]%
\centering%
\begin{subfigure}{\linewidth}%
\centering%
\includegraphics[scale=\evalgraphscale]{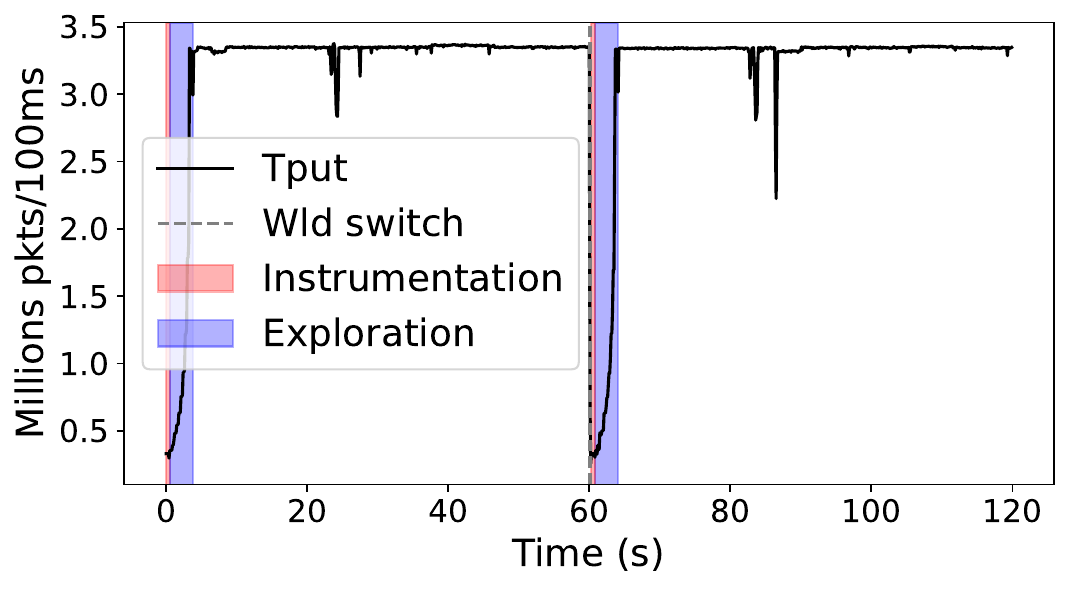}%
\end{subfigure}%

\begin{subfigure}{\linewidth}%
\centering%
\includegraphics[scale=\evalgraphscale]{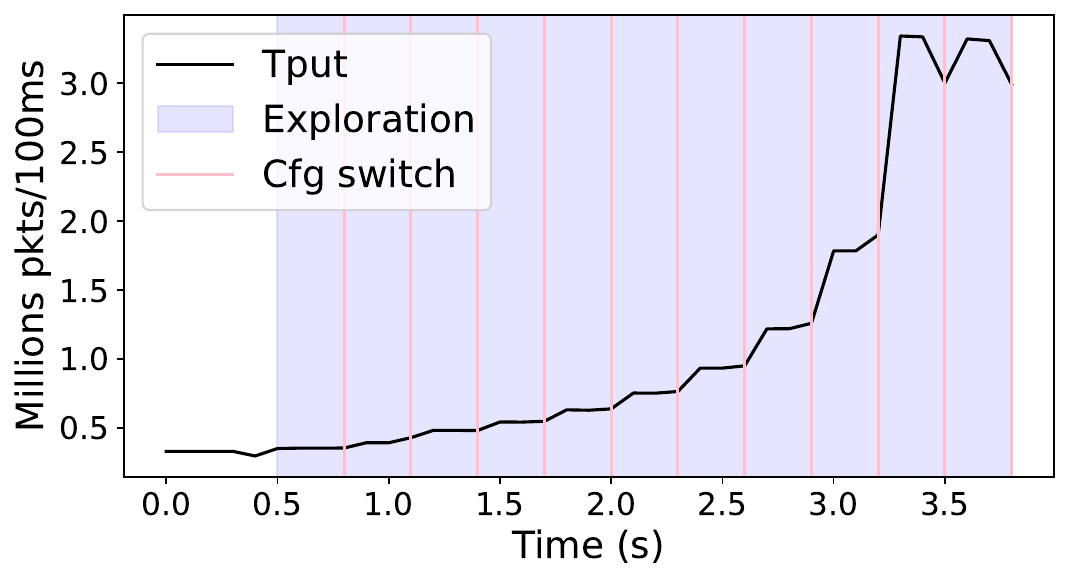}%
\end{subfigure}%
\caption{Optimal fast path size exploration with \sys.}%
\label{fig:fastclick_adaptation}%
\end{figure}

\fakepara{\sys transparently enables exploration-based adaptation with custom specializations.}
We show runtime design exploration for FastClick LinearIPLookup.
We run the FastPath router on one machine, treat that machine as the
Device Under Test and run Pktgen to execute an open-loop workload
generating packets with destination ip addresses from a given set of IP addresses. At the 1 minute-mark, we completely switch the destination IP address set with no overlap with the initial address.
The router is configured with a \sys specialization policy that triggers an exploration whenever it detects a large change ($\ge25\%$) in measured throughput.
\autoref{fig:fastclick_adaptation} shows how the throughput of the router changes for the different IP sets. 
At the start, the specialization policy triggers an exploration as it detects a large in-flux of packets, a significant change from 0 incoming packets. First, \sys runs a quick instrumentation phase for around 100ms to find the most popular incoming destination IP addresses. After the instrumentation phase, \sys switches to an exploration phase to explore the size of the fast-path, \ie, how many addresses should be included in the generated if-else source code. Once the exploration finishes, \sys switches to the best performing fast-path size and generates the specialized code which persists until the workload switch at the 1-minute mark. Due to the workload change, the specialization policy kicks in automatically, and \sys restarts the instrumentation and exploration phase. As shown in \autoref{fig:fastclick_adaptation}, \sys can quickly adapt to changes in workload to generate optimal specialized code under dynamic conditions.

\fakepara{\sys can explore large specialization spaces.} We explore the large specialization space of CloverLeaf with OPPerTune's exploration algorithm
integrated with \sys. We limit the number of configurations to be explored to 20 out of a total of $7^{24}$ total configurations. We observe that,
\sys can find an configuration that reduces the number of cycles spent per iteration of the main CloverLeaf loop by $3\times$ on average.
By leveraging state-of-the-art auto-tuning exploration strategies, \sys can efficiently find an optimal configuration from large specialization
spaces.

\subsection{\sys but at what Cost?}
\label{sec:cost}

\fakepara{Compilation cost.} \autoref{tab:jit_compilation} shows the JIT compilation time for each of the target systems. Note that, this compilation happens off the critical path so it is not a performance bottleneck. However, this cost does dictate the propagation delay, \ie, the time
taken for the specialized version of the code to be available for the execution.

\begin{table}[t]
\centering
\small
\begin{tabular}{lcc}
\toprule
\textbf{System} & \textbf{Time (ms)} & \textbf{Size (KB)} \\
\midrule
MMulBlockBench & 10 $\pm$ 1 & 10 \\
LibLPM-FP & 72 $\pm$ 9 & 80\\
Network Functions & 98 $\pm$ 5 & 67 \\
TAS & 340 $\pm$ 5 & 360 \\
FastClick & 11 $\pm$ 1 & 9.2 \\
\bottomrule
\end{tabular}
\caption{JIT compilation time and corresponding size}
\label{tab:jit_compilation}
\end{table}

\begin{figure}[t]%
\centering%
\includegraphics[scale=\evalgraphscale]{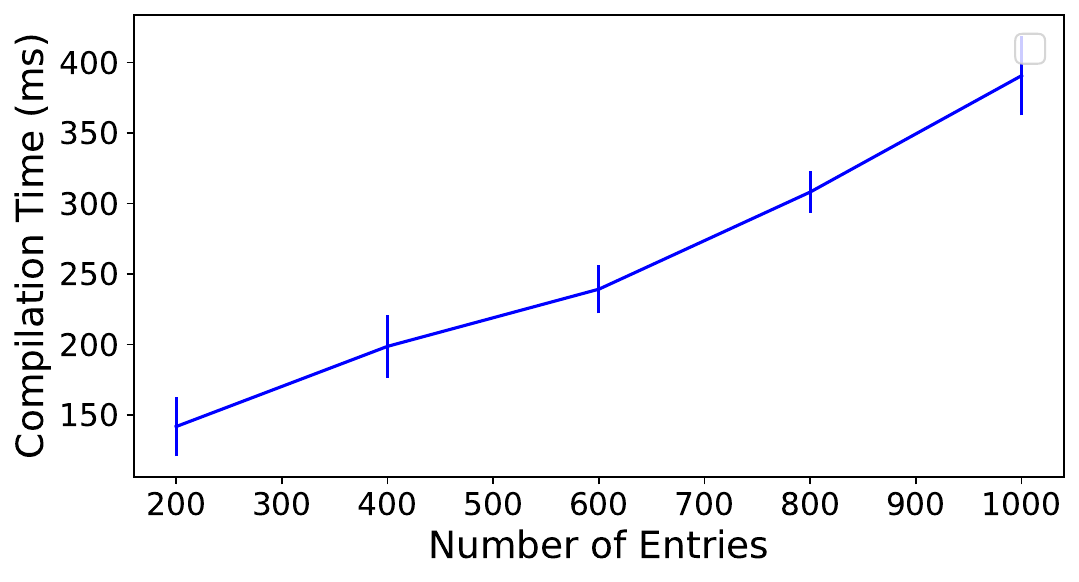}%
\caption{Compilation time of LibLPM-NI (nested-if specialization) as a function of the number of elements in the LPM lookup table.}%
\label{fig:lpm_compilation}%
\end{figure}

To analyze how the compilation time changes with the increase in the size of the JIT, we consider the code generated by \sys for the LibLPM-NI specialization.
 \autoref{fig:lpm_compilation} shows the change in the compilation time for the LibLPM-NI specialization as a function of the number
of the elements in the lpm lookup table. The compilation time increases linearly with increase in the number of lpm entries.
This is because \sys generates one basic block in the specialized code for each lpm entry. As a result, the code size generated
by \sys grows linearly with the total number of entries.

\fakepara{Developer cost.} \autoref{tab:loc} shows the necessary lines of code changes to integrate \sys specializations into the target systems. For all cases, fewer than 100 lines of code change. Implementing the custom LPM specialization only required 162 lines of code for enabling the custom user-defined LPM specialization.

\begin{table}[t]
  \centering
  \small
  \begin{tabular}{lcc}
  \toprule
  \textbf{System} & \makecell{\textbf{Handler} \\\textbf{Annotations (LoC)}} & \makecell{\textbf{Spec Policy} \\ \textbf{(LoC)}} \\
  \midrule
  MMulBlockBench & 2 & 47\\
  LibLPM & 4 & 9\\
  Network Functions & 5 & 43\\
  TAS & 8 & 38\\
  FastClick & 5 & 88\\
  \bottomrule
  \end{tabular}
  \caption{Lines of Code change required to integrate \sys.}
  \label{tab:loc}
\end{table}

\begin{figure}[t]%
\centering%
\includegraphics[scale=\evalgraphscale]{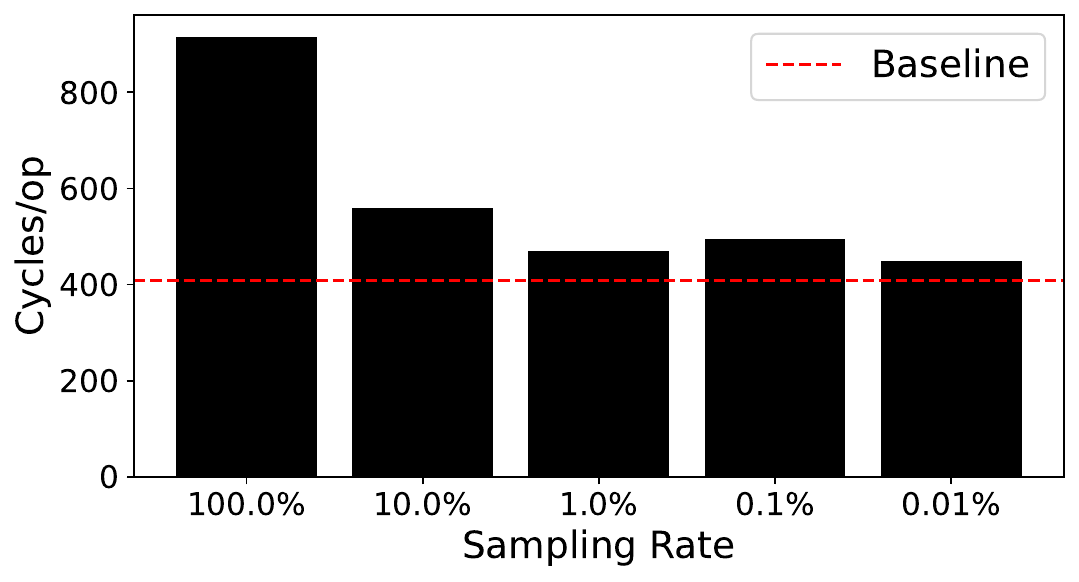}%
\caption{Impact of sampling rate on instrumentation overhead for
  LibLPM-FP configuration for Workload B.}%
\label{fig:lpm_instrumentation}%
\end{figure}

\fakepara{Instrumentation overhead.} To measure the overhead of adding instrumentation, we devise a microbenchmark called SimpleBench,
consisting of two very simple functions, \texttt{f} and \texttt{g}. \texttt{f} computes
the square of the input and \texttt{g} computes the product of its two inputs. For \texttt{f}, we add a general specialization point for its input \texttt{a}.
For \texttt{g}, we add a range-based specialization for its second input \texttt{b}.
For both the specialization points \texttt{a} and \texttt{b}, we first execute their respective functions, \texttt{f} and \texttt{g}, in normal mode and then use \sys to
add instrumentation. We execute each configuration one million times. As a baseline, each function on average takes up to 8 cycles per execution in normal mode.
For \texttt{a}, which is a general specialization point, adding instrumentation adds an overhead of around 450 to 500 cycles per execution.
For \texttt{b}, which is a specialization point designed for more efficient instrumentation for specialization points with values in fixed ranges, adding
instrumentation adds an overhead of around 1 extra cycle per operation.
When possible, \sys users should opt to use the more efficient specialization point. However, as this ma not always be possible
and users may be forced to use the general specialization point, \sys allows users to specify a sampling rate to avoid
incurring the instrumentation cost for low-latency computations. \autoref{fig:lpm_instrumentation} shows how decreasing the sampling
rate for the LibLPM-FP Workload B configuration can decrease the instrumentation overhead. 

\fakepara{Specialization guards and failures.} To measure the cost of specialization guards and failures, we use the SimpleBench microbenchmark.
For function \texttt{f}, we use \sys to specialize the value of \texttt{a} and generate two different versions. In the first version,
we disable the specialization check at the entrypoint of \texttt{f}. We then execute this version of the function with the specialized value as input one million times. 
Each execution of this version takes 7 cycles. In the second version, we enable the specialization check at the entrypoint of \texttt{f}. 
If the specialization check fails, \ie the input value of \texttt{a} does not match the specialized value of \texttt{a}, then this version of \texttt{f}
throws an exception that is caught by \sys-inserted trampoline code which subsequently calls the generic version of the function. We first execute this
version of the function with inputs such that the specialization guards always pass. This check adds an extra cycle per execution of the function.
Next, we execute this version of the function with inputs such that the specialization guards always fail. This results in exception handling behavior
which adds approximately 5000 cycles of overhead per execution. Thus, the raw cost of a specialization check is 1 extra cycle whereas that of
a specialization failure is approximately 5000 cycles.

\section{Related Work}%
\label{sec:related}

\subsection{Offline Automatic Specializations}
We classify offline automatic specialization
techniques in four categories.
(i) Compile pass selection techniques automatically select the optimal sequence of compiler optimization passes
through auto-tuning methods for trying out different combinations~\cite{park2022srtuner,pan2025synergy,pan2025hybrid,pan2025synergy},
(ii) Specialized compiler optimization passes that can generate specialized code for specific hardware features
like vectorization~\cite{naser2025vectron} or prefetching~\cite{sioutas2018loop}, or for other specific optimizations~\cite{leopoldseder2018dominance,mccandless2012compiler,rus2002hybrid,alappat2021yasksite},
(iii) Feedback-Driven Optimizations (FDO)~\cite{chen2016autofdo,panchenko2019bolt,holzle1994optimizing,dehnert2003transmeta,ottoni2018hhvm,selakovic2017actionable,stephenson2021pgz,roy2016structslim,yang2002efficient,he2024revamping,zhao2005feedback,jamilan2022apt,zhang2024rpg2,he2024revamping}, also known as PGO (Profile-Guided Optimizations),
are compiler optimizations consisting of multiple executions of the program to be optimized
by collecting relevant metrics from a benchmarking run and then using these metrics to generate optimized versions.
FDO techniques in their current form lack flexibility
as they are specialized to work well for only one workload (the workload it was trained on), 
(iv) Specialized Code Tuning approaches either use trial runs to tune specific parameters in the code~\cite{lee2019white,popov2017piecewise}
or apply specialized optimizations to different pieces of the code~\cite{teixeira2019locus,popov2017piecewise}.
Unlike \sys, existing offline automatic optimizations cannot adapt to changes in workload or environmental settings dynamically.
Consequently, these techniques are sub-optimal for the conditions they were not trained on.

\subsection{Online Automatic Optimizations}
We classify existing online automatic optimizations in four categories --- (i) runtime code selection, (ii) configuration optimizations, (iii) generic code specializations, (iv) domain-specific specializations.

\fakepara{Runtime code selection.} In this optimization technique, developers pre-generate multiple different versions of the code at compile-time
and then use measured runtime statistics and offline-trained models to switch between the versions~\cite{kambadur2016nrg,srinivas2015reactive,sriraman2018mutune,costa2018collectionswitch}. These techniques are limited as they must anticipate and generate all possible versions a priori.
Consequently, in contrast to \sys, these techniques can not fully leverage runtime information for specialization.

\fakepara{Config optimizations.} Runtime auto-tuning techniques such as OtterTune~\cite{van2017automatic} often perform tuning through trial sessions
to find optimal values for different configuration knobs.
Some AutoTuning techniques such as OPPerTune~\cite{somashekar2024oppertune}
and SoftSKU~\cite{sriraman2019softsku} apply the AutoTuning paradigm of measurement-driven search at runtime without any trials.
However, these techniques are critically limited to environmental parameters and knobs as they work transparently with respect to the deployed application.
These techniques are currently not applicable to code optimizations or specializations.
\sys enables the auto-tuning paradigm of measurement-driven search at runtime to find the best combination of code specializations for the system.

\fakepara{Code specializations.} Runtime code specialization techniques~\cite{consel1996general,consel2005uniform,grant1999evaluation,dean1995selective,arnold2000adaptive,prokopec2019optimization} produce more efficient versions of the code by exploiting values and invariants that
only exist at runtime. 
Today, these specializations manifest in three ways: (i) binary rewriting techniques~\cite{ali2017dynamic}; 
(ii) Transparent Dynamic Optimization (TDO) techniques such as Dynamo~\cite{bala2000dynamo}, The Transmeta Code Morphing Software~\cite{dehnert2003transmeta}, DynamoRIO~\cite{bruening2003infrastructure}. These optimize code at runtime without requiring any modifications
by capturing and optimizing traces (sequences of instructions) that are commonly executed;
(iii) optimizations in JIT compilers for interpreted languages in the form of Type Specialization~\cite{chevalier2010optimizing,ottoni2018hhvm,chevalier2015interprocedural} for specializing the types of data for dynamically typed languages or as Value Specialization for interpreted languages~\cite{de2014just,lima2020guided} where the parameter values of hot functions are converted into constants.
These specializations may be optionally applied depending on the computation context~\cite{jain2016phase,poesia2020dynamic,fluckiger2020contextual,mitra2017phase}.
In contrast to \sys, they are generic and typically done automatically by the runtime based on internal cost models
which may not be appropriate for specific application contexts. Moreover, these specializations are limited in scope and
do not take the system's end-to-end performance into consideration
and do not leverage domain- and situation-specific optimizations.

\fakepara{Domain-specific specializations.} Recently, value specializations have extended beyond interpreted languages for specific use cases such as value specializations for GPU kernels~\cite{georgakoudis2025proteus}, or through incremental specializations for packet processing frameworks~\cite{ruffy2024incremental,miano2022domain,farshin2021packetmill}. \sys provides a general framework for configuring, applying, exploring,
and selecting such specializations at runtime. 

\section{Discussion}
\label{sec:discussion}

\fakepara{Debugging.} \sys makes it harder for developers to debug performance issues.
This is because the generated specialized code is often different from the code
that was originally written by the developer, which makes it harder for the developers
to fully understand the execution sequences. This is further exacerbated by the fact that
different workloads may lead to different specializations making it difficult
to reproduce issues seen in production. Combining \sys with monitoring
and distributed tracing techniques could offer developers a solution
to their debugging problems.

\fakepara{Leveraging scale \& ML for exploration.}
Several ML-based techniques exist for tuning configurations~\cite{van2017automatic,akgun2023improving,cereda2021cgptuner,li2018metis,lin2022adaptive,mahgoub2020optimuscloud,somashekar2022reducing,somashekar2024oppertune,duan2009tuning,jamilan2022apt,swaminathan2017off,ruffy2024incremental}. 
These ML techniques can be implemented as algorithms with
OPPerTune~\cite{somashekar2024oppertune}, which we have already integrated with \sys. Moreover, \sys could further incorporate SmartChoices~\cite{carbune2018smartchoices} to allow
different ML techniques to apply to specific specialization points.
As part of future work, \sys could potentially also parallelize
exploration across a large number of replicas to explore larger
subsets of the specialization space.

\fakepara{Limitations.} While \sys is designed to operate in dynamic and irregular operating conditions,
exploration overhead overshadows performance benefits if the operating conditions do not remain stable
for long enough to reap the benefits of the specialization optimizations.
Ideally, developers should configure the specialization policy to opt
for the general version of the handler code in such scenarios.

\section{Conclusions}

In this work, we presented \sys, a framework enabling online workload-driven runtime specialization of systems to improve performance.
\sys provides a toolkit for developers to implement system-specific specializations that
utilize runtime data and invariants to generate more performant systems. \sys provides the necessary tooling
for developers to configure automatic runtime exploration of the space of potential specializations
to find the best performing configuration at each instant. By combining developer insight with automated compiler optimizations, \sys enables systems to adapt efficiently to 
changing workloads and environments with minimal manual effort. \if \ANON 0
\section*{Acknowledgments}
We thank Yuchen Qian for his contributions to an early prototype Iridescent implementation.
 \fi

\bibliographystyle{plain}
\bibliography{paper,bibdb/papers,bibdb/strings,bibdb/defs}

\begin{thebibliography}{10}

\bibitem{akgun2023improving}
Ibrahim~Umit Akgun, Ali~Selman Aydin, Andrew Burford, Michael McNeill, Michael
  Arkhangelskiy, and Erez Zadok.
\newblock Improving storage systems using machine learning.
\newblock {\em ACM Transactions on Storage}, 19(1):1--30, 2023.

\bibitem{alappat2021yasksite}
Christie~L Alappat, Johannes Seiferth, Georg Hager, Matthias Korch, Thomas
  Rauber, and Gerhard Wellein.
\newblock Yasksite: stencil optimization techniques applied to explicit ode
  methods on modern architectures.
\newblock In {\em 2021 IEEE/ACM International Symposium on Code Generation and
  Optimization (CGO)}, pages 174--186. IEEE, 2021.

\bibitem{ali2017dynamic}
AP~Arif Ali and Erven Rohou.
\newblock Dynamic function specialization.
\newblock In {\em 2017 International Conference on Embedded Computer Systems:
  Architectures, Modeling, and Simulation (SAMOS)}, pages 163--170. IEEE, 2017.

\bibitem{alipourfard2017cherrypick}
Omid Alipourfard, Hongqiang~Harry Liu, Jianshu Chen, Shivaram Venkataraman,
  Minlan Yu, and Ming Zhang.
\newblock {CherryPick}: Adaptively unearthing the best cloud configurations for
  big data analytics.
\newblock In {\em 14th USENIX Symposium on Networked Systems Design and
  Implementation (NSDI 17)}, pages 469--482, 2017.

\bibitem{arnold2000adaptive}
Matthew Arnold, Stephen Fink, David Grove, Michael Hind, and Peter~F Sweeney.
\newblock Adaptive optimization in the jalapeno jvm.
\newblock In {\em Proceedings of the 15th ACM SIGPLAN conference on
  Object-oriented programming, systems, languages, and applications}, pages
  47--65, 2000.

\bibitem{bala2000dynamo}
Vasanth Bala, Evelyn Duesterwald, and Sanjeev Banerjia.
\newblock Dynamo: A transparent dynamic optimization system.
\newblock In {\em Proceedings of the ACM SIGPLAN 2000 conference on Programming
  language design and implementation}, pages 1--12, 2000.

\bibitem{barbette2015fast}
Tom Barbette, Cyril Soldani, and Laurent Mathy.
\newblock Fast userspace packet processing.
\newblock In {\em 2015 ACM/IEEE Symposium on Architectures for Networking and
  Communications Systems (ANCS)}, pages 5--16. IEEE, 2015.

\bibitem{bhatia2004automatic}
Sapan Bhatia, Charles Consel, A-F Le~Meur, and Calton Pu.
\newblock Automatic specialization of protocol stacks in operating system
  kernels.
\newblock In {\em 29th Annual IEEE International Conference on Local Computer
  Networks}, pages 152--159. IEEE, 2004.

\bibitem{bruening2003infrastructure}
Derek Bruening, Timothy Garnett, and Saman Amarasinghe.
\newblock An infrastructure for adaptive dynamic optimization.
\newblock In {\em International Symposium on Code Generation and Optimization,
  2003. CGO 2003.}, pages 265--275. IEEE, 2003.

\bibitem{carbune2018smartchoices}
Victor Carbune, Thierry Coppey, Alexander Daryin, Thomas Deselaers, Nikhil
  Sarda, and Jay Yagnik.
\newblock Smartchoices: hybridizing programming and machine learning.
\newblock {\em arXiv preprint arXiv:1810.00619}, 2018.

\bibitem{carr1989blocking}
Steve Carr and Ken Kennedy.
\newblock Blocking linear algebra codes for memory hierarchies.
\newblock In {\em PPSC}, pages 400--405. Citeseer, 1989.

\bibitem{cereda2021cgptuner}
Stefano Cereda, Stefano Valladares, Paolo Cremonesi, and Stefano Doni.
\newblock Cgptuner: a contextual gaussian process bandit approach for the
  automatic tuning of it configurations under varying workload conditions.
\newblock {\em Proceedings of the VLDB Endowment}, 14(8):1401--1413, 2021.

\bibitem{chen2016autofdo}
Dehao Chen, David~Xinliang Li, and Tipp Moseley.
\newblock Autofdo: Automatic feedback-directed optimization for warehouse-scale
  applications.
\newblock In {\em Proceedings of the 2016 International Symposium on Code
  Generation and Optimization}, pages 12--23, 2016.

\bibitem{chevalier2015interprocedural}
Maxime Chevalier-Boisvert and Marc Feeley.
\newblock Interprocedural type specialization of javascript programs without
  type analysis.
\newblock {\em arXiv preprint arXiv:1511.02956}, 2015.

\bibitem{chevalier2010optimizing}
Maxime Chevalier-Boisvert, Laurie Hendren, and Clark Verbrugge.
\newblock Optimizing matlab through just-in-time specialization.
\newblock In {\em International Conference on Compiler Construction}, pages
  46--65. Springer, 2010.

\bibitem{choi2025ml}
Inho Choi, Anand Bonde, Jing Liu, Joshua Fried, Irene Zhang, and Jialin Li.
\newblock Ml-native dataplane operating systems.
\newblock In {\em Proceedings of the 16th ACM SIGOPS Asia-Pacific Workshop on
  Systems}, pages 8--14, 2025.

\bibitem{consel2005uniform}
Charles Consel, Luke Hornof, Fran{\c{c}}ois No{\"e}l, Jacques Noy{\'e}, and
  Nicolae Volanschi.
\newblock A uniform approach for compile-time and run-time specialization.
\newblock In {\em Partial Evaluation: International Seminar Dagstuhl Castle,
  Germany, February 12--16, 1996 Selected Papers}, pages 54--72. Springer,
  2005.

\bibitem{consel1996general}
Charles Consel and Fran{\c{c}}ois No{\"e}l.
\newblock A general approach for run-time specialization and its application to
  c.
\newblock In {\em Proceedings of the 23rd ACM SIGPLAN-SIGACT symposium on
  Principles of programming languages}, pages 145--156, 1996.

\bibitem{specaccel}
Standard Performance~Evaluation Corporation.
\newblock Spec accel benchmark suite.
\newblock Available from https://www.spec.org/accel, 2023.

\bibitem{costa2018collectionswitch}
Diego Costa and Artur Andrzejak.
\newblock Collectionswitch: A framework for efficient and dynamic collection
  selection.
\newblock In {\em Proceedings of the 2018 International Symposium on Code
  Generation and Optimization}, pages 16--26, 2018.

\bibitem{de2014just}
Igor~Rafael de~Assis~Costa, Henrique~Nazar{\'e} Santos, P{\'e}ricles~Rafael
  Alves, and Fernando Magno~Quintao Pereira.
\newblock Just-in-time value specialization.
\newblock {\em Computer Languages, Systems \& Structures}, 40(2):37--52, 2014.

\bibitem{dean1995selective}
Jeffrey Dean, Craig Chambers, and David Grove.
\newblock Selective specialization for object-oriented languages.
\newblock {\em ACM SIGPLAN Notices}, 30(6):93--102, 1995.

\bibitem{dehnert2003transmeta}
James~C Dehnert, Brian~K Grant, John~P Banning, Richard Johnson, Thomas
  Kistler, Alexander Klaiber, and Jim Mattson.
\newblock The transmeta code morphing/spl trade/software: using speculation,
  recovery, and adaptive retranslation to address real-life challenges.
\newblock In {\em International Symposium on Code Generation and Optimization,
  2003. CGO 2003.}, pages 15--24. IEEE, 2003.

\bibitem{deng2020redundant}
Bangwen Deng, Wenfei Wu, and Linhai Song.
\newblock Redundant logic elimination in network functions.
\newblock In {\em Proceedings of the Symposium on SDN Research}, pages 34--40,
  2020.

\bibitem{software:dpdk}
{DPDK Project}.
\newblock Data plane development kit.
\newblock \url{http://www.dpdk.org/}, 2022.
\newblock Retrieved Feb 2, 2022.

\bibitem{duan2009tuning}
Songyun Duan, Vamsidhar Thummala, and Shivnath Babu.
\newblock Tuning database configuration parameters with ituned.
\newblock {\em Proceedings of the VLDB Endowment}, 2(1):1246--1257, 2009.

\bibitem{farshin2021packetmill}
Alireza Farshin, Tom Barbette, Amir Roozbeh, Gerald~Q Maguire~Jr, and Dejan
  Kosti{\'c}.
\newblock Packetmill: toward per-core 100-gbps networking.
\newblock In {\em Proceedings of the 26th ACM International Conference on
  Architectural Support for Programming Languages and Operating Systems}, pages
  1--17, 2021.

\bibitem{fluckiger2020contextual}
Olivier Fl{\"u}ckiger, Guido Chari, Ming-Ho Yee, Jan Je{\v{c}}men, Jakob Hain,
  and Jan Vitek.
\newblock Contextual dispatch for function specialization.
\newblock {\em Proceedings of the ACM on Programming Languages},
  4(OOPSLA):1--24, 2020.

\bibitem{georgakoudis2025proteus}
Giorgis Georgakoudis, Konstantinos Parasyris, and David Beckingsale.
\newblock Proteus: Portable runtime optimization of gpu kernel execution with
  just-in-time compilation.
\newblock In {\em Proceedings of the 23rd ACM/IEEE International Symposium on
  Code Generation and Optimization}, pages 507--522, 2025.

\bibitem{ghasemirahni2022packet}
Hamid Ghasemirahni, Tom Barbette, Georgios~P Katsikas, Alireza Farshin, Amir
  Roozbeh, Massimo Girondi, Marco Chiesa, Gerald~Q Maguire~Jr, and Dejan
  Kosti{\'c}.
\newblock Packet order matters! improving application performance by
  deliberately delaying packets.
\newblock In {\em 19th USENIX Symposium on Networked Systems Design and
  Implementation (NSDI 22)}, pages 807--827, 2022.

\bibitem{ghasemirahni2024just}
Hamid Ghasemirahni, Alireza Farshin, Dejan Kostic, and Marco Chiesa.
\newblock Just-in-time packet state prefetching.
\newblock {\em arXiv preprint arXiv:2407.04344}, 2024.

\bibitem{grant1999evaluation}
Brian Grant, Matthai Philipose, Markus Mock, Craig Chambers, and Susan~J
  Eggers.
\newblock An evaluation of staged run-time optimizations in dyc.
\newblock {\em ACM SIGPLAN Notices}, 34(5):293--304, 1999.

\bibitem{he2024revamping}
Wenlei He, Hongtao Yu, Lei Wang, and Taewook Oh.
\newblock Revamping sampling-based pgo with context-sensitivity and
  pseudo-instrumentation.
\newblock In {\em 2024 IEEE/ACM International Symposium on Code Generation and
  Optimization (CGO)}, pages 322--333. IEEE, 2024.

\bibitem{holzle1994optimizing}
Urs H{\"o}lzle and David Ungar.
\newblock Optimizing dynamically-dispatched calls with run-time type feedback.
\newblock In {\em Proceedings of the ACM SIGPLAN 1994 conference on Programming
  language design and implementation}, pages 326--336, 1994.

\bibitem{pixartifact}
Rishabh Iyer, Katerina Argyraki, and George Candea.
\newblock Performance interface extractor (pix) artifact.
\newblock Accessed November 2024 from \url{https://github.com/dslab-epfl/pix},
  2022.

\bibitem{iyer2022performance}
Rishabh Iyer, Katerina Argyraki, and George Candea.
\newblock Performance interfaces for network functions.
\newblock In {\em 19th USENIX Symposium on Networked Systems Design and
  Implementation (NSDI 22)}, pages 567--584, 2022.

\bibitem{jain2016phase}
Era Jain and Subhajit Roy.
\newblock Phase directed compiler optimizations.
\newblock In {\em 2016 IEEE 23rd International Conference on High Performance
  Computing (HiPC)}, pages 270--279. IEEE, 2016.

\bibitem{jamilan2022apt}
Saba Jamilan, Tanvir~Ahmed Khan, Grant Ayers, Baris Kasikci, and Heiner Litz.
\newblock Apt-get: Profile-guided timely software prefetching.
\newblock In {\em Proceedings of the Seventeenth European Conference on
  Computer Systems}, pages 747--764, 2022.

\bibitem{kalia:erpc}
Anuj Kalia, Michael Kaminsky, and David Andersen.
\newblock Datacenter {RPCs} can be general and fast.
\newblock In {\em 16th USENIX Symposium on Networked Systems Design and
  Implementation}, NSDI, 2019.

\bibitem{kambadur2016nrg}
Melanie Kambadur and Martha~A Kim.
\newblock Nrg-loops: adjusting power from within applications.
\newblock In {\em Proceedings of the 2016 International Symposium on Code
  Generation and Optimization}, pages 206--215, 2016.

\bibitem{karthikeyan2023selftune}
Ajaykrishna Karthikeyan, Nagarajan Natarajan, Gagan Somashekar, Lei Zhao,
  Ranjita Bhagwan, Rodrigo Fonseca, Tatiana Racheva, and Yogesh Bansal.
\newblock {SelfTune}: Tuning cluster managers.
\newblock In {\em 20th USENIX Symposium on Networked Systems Design and
  Implementation (NSDI 23)}, pages 1097--1114, 2023.

\bibitem{kaufmann:tas}
Antoine Kaufmann, Tim Stamler, Simon Peter, Naveen~Kr. Sharma, Arvind
  Krishnamurthy, and Thomas Anderson.
\newblock {TAS}: {TCP} acceleration as an {OS} service.
\newblock In {\em 14th ACM European Conference on Computer Systems}, EuroSys,
  2019.

\bibitem{kohler2002clickpl}
Eddie Kohler, Robert Morris, and Benjie Chen.
\newblock Programming language optimizations for modular router configurations.
\newblock In {\em Proceedings of the 10th International Conference on
  Architectural Support for Programming Languages and Operating Systems},
  ASPLOS X, pages 251--263, New York, NY, USA, 2002. Association for Computing
  Machinery.

\bibitem{kohler:click}
Eddie Kohler, Robert Morris, Benjie Chen, John Jannotti, and M.~Frans Kaashoek.
\newblock The {Click} modular router.
\newblock {\em ACM Transactions on Computer Systems}, 18(3):263--297, August
  2000.

\bibitem{lattner:lllvm}
Chris Lattner and Vikram Adve.
\newblock {LLVM}: A compilation framework for lifelong program analysis \&
  transformation.
\newblock In {\em 9th International Symposium on Code Generation and
  Optimization}, CGO, 2004.

\bibitem{lee2019white}
Wen-Chuan Lee, Yingqi Liu, Peng Liu, Shiqing Ma, Hongjun Choi, Xiangyu Zhang,
  and Rajiv Gupta.
\newblock White-box program tuning.
\newblock In {\em 2019 IEEE/ACM International Symposium on Code Generation and
  Optimization (CGO)}, pages 122--135. IEEE, 2019.

\bibitem{leopoldseder2018dominance}
David Leopoldseder, Lukas Stadler, Thomas W{\"u}rthinger, Josef Eisl, Doug
  Simon, and Hanspeter M{\"o}ssenb{\"o}ck.
\newblock Dominance-based duplication simulation (dbds): code duplication to
  enable compiler optimizations.
\newblock In {\em Proceedings of the 2018 International Symposium on Code
  Generation and Optimization}, pages 126--137, 2018.

\bibitem{li2018metis}
Zhao~Lucis Li, Chieh-Jan~Mike Liang, Wenjia He, Lianjie Zhu, Wenjun Dai, Jin
  Jiang, and Guangzhong Sun.
\newblock Metis: Robustly tuning tail latencies of cloud systems.
\newblock In {\em 2018 USENIX Annual Technical Conference (USENIX ATC 18)},
  pages 981--992, 2018.

\bibitem{lima2020guided}
Caio Lima, Junio Cezar, Guilherme~Vieira Leobas, Erven Rohou, and Fernando
  Magno~Quint{\~a}o Pereira.
\newblock Guided just-in-time specialization.
\newblock {\em Science of Computer Programming}, 185:102318, 2020.

\bibitem{lin2022adaptive}
Chen Lin, Junqing Zhuang, Jiadong Feng, Hui Li, Xuanhe Zhou, and Guoliang Li.
\newblock Adaptive code learning for spark configuration tuning.
\newblock In {\em 2022 IEEE 38th International Conference on Data Engineering
  (ICDE)}, pages 1995--2007. IEEE, 2022.

\bibitem{mahgoub2020optimuscloud}
Ashraf Mahgoub, Alexander~Michaelson Medoff, Rakesh Kumar, Subrata Mitra, Ana
  Klimovic, Somali Chaterji, and Saurabh Bagchi.
\newblock {OPTIMUSCLOUD}: Heterogeneous configuration optimization for
  distributed databases in the cloud.
\newblock In {\em 2020 USENIX Annual Technical Conference (USENIX ATC 20)},
  pages 189--203, 2020.

\bibitem{mccandless2012compiler}
Jason Mccandless and David Gregg.
\newblock Compiler techniques to improve dynamic branch prediction for indirect
  jump and call instructions.
\newblock {\em ACM Transactions on Architecture and Code Optimization (TACO)},
  8(4):1--20, 2012.

\bibitem{miano2022domain}
Sebastiano Miano, Alireza Sanaee, Fulvio Risso, G{\'a}bor R{\'e}tv{\'a}ri, and
  Gianni Antichi.
\newblock Domain specific run time optimization for software data planes.
\newblock In {\em Proceedings of the 27th ACM International Conference on
  Architectural Support for Programming Languages and Operating Systems}, pages
  1148--1164, 2022.

\bibitem{mitra2017phase}
Subrata Mitra, Manish~K Gupta, Sasa Misailovic, and Saurabh Bagchi.
\newblock Phase-aware optimization in approximate computing.
\newblock In {\em 2017 IEEE/ACM International Symposium on Code Generation and
  Optimization (CGO)}, pages 185--196. IEEE, 2017.

\bibitem{molnar2016dataplane}
L{\'a}szl{\'o} Moln{\'a}r, Gergely Pongr{\'a}cz, G{\'a}bor Enyedi,
  Zolt{\'a}n~Lajos Kis, Levente Csikor, Ferenc Juh{\'a}sz, Attila
  K{\H{o}}r{\"o}si, and G{\'a}bor R{\'e}tv{\'a}ri.
\newblock Dataplane specialization for high-performance openflow software
  switching.
\newblock In {\em Proceedings of the 2016 ACM SIGCOMM Conference}, pages
  539--552, 2016.

\bibitem{naser2025vectron}
Sourena Naser~Moghaddasi, Haris Smajlovi{\'c}, Ariya Shajii, and Ibrahim
  Numanagi{\'c}.
\newblock Vectron: A dynamic programming auto-vectorization framework.
\newblock In {\em Proceedings of the 23rd ACM/IEEE International Symposium on
  Code Generation and Optimization}, pages 644--659, 2025.

\bibitem{ottoni2018hhvm}
Guilherme Ottoni.
\newblock Hhvm jit: A profile-guided, region-based compiler for php and hack.
\newblock In {\em Proceedings of the 39th ACM SIGPLAN Conference on Programming
  Language Design and Implementation}, pages 151--165, 2018.

\bibitem{pan2025synergy}
Haolin Pan, Jinyuan Dong, Mingjie Xing, and Yanjun Wu.
\newblock Synergy-guided compiler auto-tuning of nested llvm pass pipelines.
\newblock {\em arXiv preprint arXiv:2510.13184}, 2025.

\bibitem{pan2025hybrid}
Haolin Pan, Hongbin Zhang, Mingjie Xing, and Yanjun Wu.
\newblock A hybrid, knowledge-guided evolutionary framework for personalized
  compiler auto-tuning.
\newblock {\em arXiv preprint arXiv:2510.14292}, 2025.

\bibitem{pan2024hoda}
Heng Pan, Peng He, Zhenyu Li, Pan Zhang, Junjie Wan, Yuhao Zhou, XiongChun
  Duan, Yu~Zhang, and Gaogang Xie.
\newblock Hoda: a high-performance open vswitch dataplane with multiple
  specialized data paths.
\newblock In {\em Proceedings of the Nineteenth European Conference on Computer
  Systems}, pages 82--98, 2024.

\bibitem{panchenko2019bolt}
Maksim Panchenko, Rafael Auler, Bill Nell, and Guilherme Ottoni.
\newblock Bolt: a practical binary optimizer for data centers and beyond.
\newblock In {\em 2019 IEEE/ACM International Symposium on Code Generation and
  Optimization (CGO)}, pages 2--14. IEEE, 2019.

\bibitem{park2022srtuner}
Sunghyun Park, Salar Latifi, Yongjun Park, Armand Behroozi, Byungsoo Jeon, and
  Scott Mahlke.
\newblock Srtuner: Effective compiler optimization customization by exposing
  synergistic relations.
\newblock In {\em 2022 IEEE/ACM International Symposium on Code Generation and
  Optimization (CGO)}, pages 118--130. IEEE, 2022.

\bibitem{poesia2020dynamic}
Gabriel Poesia and Fernando Magno~Quint{\~a}o Pereira.
\newblock Dynamic dispatch of context-sensitive optimizations.
\newblock {\em Proceedings of the ACM on Programming Languages},
  4(OOPSLA):1--28, 2020.

\bibitem{popov2017piecewise}
Mihail Popov, Chadi Akel, Yohan Chatelain, William Jalby, and Pablo
  de~Oliveira~Castro.
\newblock Piecewise holistic autotuning of parallel programs with cere.
\newblock {\em Concurrency and Computation: Practice and Experience},
  29(15):e4190, 2017.

\bibitem{prokopec2019optimization}
Aleksandar Prokopec, Gilles Duboscq, David Leopoldseder, and Thomas
  W{\"\i}rthinger.
\newblock An optimization-driven incremental inline substitution algorithm for
  just-in-time compilers.
\newblock In {\em 2019 IEEE/ACM International Symposium on Code Generation and
  Optimization (CGO)}, pages 164--179. IEEE, 2019.

\bibitem{liblpm}
Mindaugas Rasiuckevicius.
\newblock Longest prefix match (lpm) library.
\newblock Accessed June 2025 from \url{https://github.com/rmind/liblpm}, 2022.

\bibitem{roy2016structslim}
Probir Roy and Xu~Liu.
\newblock Structslim: A lightweight profiler to guide structure splitting.
\newblock In {\em Proceedings of the 2016 International Symposium on Code
  Generation and Optimization}, pages 36--46, 2016.

\bibitem{ruffy2024incremental}
Fabian Ruffy, Zhanghan Wang, Gianni Antichi, Aurojit Panda, and Anirudh
  Sivaraman.
\newblock Incremental specialization of network programs.
\newblock In {\em Proceedings of the 23rd ACM Workshop on Hot Topics in
  Networks}, pages 264--272, 2024.

\bibitem{rus2002hybrid}
Silvius Rus, Lawrence Rauchwerger, and Jay Hoeflinger.
\newblock Hybrid analysis: static \& dynamic memory reference analysis.
\newblock In {\em Proceedings of the 16th international conference on
  Supercomputing}, pages 274--284, 2002.

\bibitem{rzadca2020autopilot}
Krzysztof Rzadca, Pawel Findeisen, Jacek Swiderski, Przemyslaw Zych, Przemyslaw
  Broniek, Jarek Kusmierek, Pawel Nowak, Beata Strack, Piotr Witusowski, Steven
  Hand, et~al.
\newblock Autopilot: workload autoscaling at google.
\newblock In {\em Proceedings of the Fifteenth European Conference on Computer
  Systems}, pages 1--16, 2020.

\bibitem{selakovic2017actionable}
Marija Selakovic, Thomas Glaser, and Michael Pradel.
\newblock An actionable performance profiler for optimizing the order of
  evaluations.
\newblock In {\em Proceedings of the 26th ACM SIGSOFT International Symposium
  on Software Testing and Analysis}, pages 170--180, 2017.

\bibitem{sharif2018trimmer}
Hashim Sharif, Muhammad Abubakar, Ashish Gehani, and Fareed Zaffar.
\newblock Trimmer: application specialization for code debloating.
\newblock In {\em Proceedings of the 33rd ACM/IEEE International Conference on
  Automated Software Engineering}, pages 329--339, 2018.

\bibitem{sioutas2018loop}
Savvas Sioutas, Sander Stuijk, Henk Corporaal, Twan Basten, and Lou Somers.
\newblock Loop transformations leveraging hardware prefetching.
\newblock In {\em Proceedings of the 2018 International Symposium on Code
  Generation and Optimization}, pages 254--264, 2018.

\bibitem{somashekar2022reducing}
Gagan Somashekar, Amoghavarsha Suresh, Saurabh Tyagi, Vikas Dhyani, K~Donkada,
  Anurag Pradhan, and Anshul Gandhi.
\newblock Reducing the tail latency of microservices applications via optimal
  configuration tuning.
\newblock In {\em 2022 IEEE International Conference on Autonomic Computing and
  Self-Organizing Systems (ACSOS)}, pages 111--120. IEEE, 2022.

\bibitem{somashekar2024oppertune}
Gagan Somashekar, Karan Tandon, Anush Kini, Chieh-Chun Chang, Petr Husak,
  Ranjita Bhagwan, Mayukh Das, Anshul Gandhi, and Nagarajan Natarajan.
\newblock {OPPerTune}: Post-deployment configuration tuning of services made
  easy.
\newblock In {\em 21st USENIX Symposium on Networked Systems Design and
  Implementation (NSDI 24)}, pages 1101--1120, 2024.

\bibitem{srinivas2015reactive}
Jithendra Srinivas, Wei Ding, and Mahmut Kandemir.
\newblock Reactive tiling.
\newblock In {\em 2015 IEEE/ACM International Symposium on Code Generation and
  Optimization (CGO)}, pages 91--102. IEEE, 2015.

\bibitem{sriraman2019softsku}
Akshitha Sriraman, Abhishek Dhanotia, and Thomas~F Wenisch.
\newblock Softsku: Optimizing server architectures for microservice diversity@
  scale.
\newblock In {\em Proceedings of the 46th International Symposium on Computer
  Architecture}, pages 513--526, 2019.

\bibitem{sriraman2018mutune}
Akshitha Sriraman and Thomas~F Wenisch.
\newblock {$\mu$Tune}: Auto-tuned threading for {OLDI} microservices.
\newblock In {\em 13th USENIX Symposium on Operating Systems Design and
  Implementation (OSDI 18)}, pages 177--194, 2018.

\bibitem{stephenson2021pgz}
Mark Stephenson and Ram Rangan.
\newblock Pgz: automatic zero-value code specialization.
\newblock In {\em Proceedings of the 30th ACM SIGPLAN International Conference
  on Compiler Construction}, pages 36--46, 2021.

\bibitem{swaminathan2017off}
Adith Swaminathan, Akshay Krishnamurthy, Alekh Agarwal, Miro Dudik, John
  Langford, Damien Jose, and Imed Zitouni.
\newblock Off-policy evaluation for slate recommendation.
\newblock {\em Advances in Neural Information Processing Systems}, 30, 2017.

\bibitem{teixeira2019locus}
SFX~Thiago Teixeira, Corinne Ancourt, David Padua, and William Gropp.
\newblock Locus: a system and a language for program optimization.
\newblock In {\em 2019 IEEE/ACM International Symposium on Code Generation and
  Optimization (CGO)}, pages 217--228. IEEE, 2019.

\bibitem{van2017automatic}
Dana Van~Aken, Andrew Pavlo, Geoffrey~J Gordon, and Bohan Zhang.
\newblock Automatic database management system tuning through large-scale
  machine learning.
\newblock In {\em Proceedings of the 2017 ACM international conference on
  management of data}, pages 1009--1024, 2017.

\bibitem{wintermeyer2020p2go}
Patrick Wintermeyer, Maria Apostolaki, Alexander Dietm{\"u}ller, and Laurent
  Vanbever.
\newblock P2go: P4 profile-guided optimizations.
\newblock In {\em Proceedings of the 19th ACM Workshop on Hot Topics in
  Networks}, pages 146--152, 2020.

\bibitem{wolfe1987iteration}
Michael Wolfe.
\newblock Iteration space tiling for memory hierarchies.
\newblock In {\em Proceedings of the Third SIAM Conference on Parallel
  Processing for Scientific Computing}, pages 357--361, 1987.

\bibitem{wolfe1989more}
Michael Wolfe.
\newblock More iteration space tiling.
\newblock In {\em Proceedings of the 1989 ACM/IEEE conference on
  Supercomputing}, pages 655--664, 1989.

\bibitem{wu2024tomur}
Shaofeng Wu, Qiang Su, Zhixiong Niu, and Hong Xu.
\newblock Tomur: Traffic-aware performance prediction of on-nic network
  functions with multi-resource contention.
\newblock {\em arXiv preprint arXiv:2405.05529}, 2024.

\bibitem{yang2002efficient}
Minghui Yang, Gang-Ryung Uh, and David~B Whalley.
\newblock Efficient and effective branch reordering using profile data.
\newblock {\em ACM Transactions on Programming Languages and Systems (TOPLAS)},
  24(6):667--697, 2002.

\bibitem{zaostrovnykh2019verifying}
Arseniy Zaostrovnykh, Solal Pirelli, Rishabh Iyer, Matteo Rizzo, Luis Pedrosa,
  Katerina Argyraki, and George Candea.
\newblock Verifying software network functions with no verification expertise.
\newblock In {\em Proceedings of the 27th ACM Symposium on Operating Systems
  Principles}, pages 275--290, 2019.

\bibitem{zhang2024rpg2}
Yuxuan Zhang, Nathan Sobotka, Soyoon Park, Saba Jamilan, Tanvir~Ahmed Khan,
  Baris Kasikci, Gilles~A Pokam, Heiner Litz, and Joseph Devietti.
\newblock Rpg2: Robust profile-guided runtime prefetch generation.
\newblock In {\em Proceedings of the 29th ACM International Conference on
  Architectural Support for Programming Languages and Operating Systems, Volume
  2}, pages 999--1013, 2024.

\bibitem{zhao2005feedback}
Peng Zhao and Jos{\'e}~Nelson Amaral.
\newblock Feedback-directed switch-case statement optimization.
\newblock In {\em 2005 International Conference on Parallel Processing
  Workshops (ICPPW'05)}, pages 295--302. IEEE, 2005.

\end{thebibliography}

\label{page:last}
\end{document}